\documentclass[a4paper,11pt]{article}
\pdfoutput=1

\usepackage{jcappub}

\usepackage{amsmath}
\usepackage{textcomp}
\usepackage{array}
\usepackage{graphicx}

\newcommand{\Fluka}{F{\sc{luka}} }
\newcommand{\Flukadot}{F{\sc{luka}}}
\newcommand{\Geant}{G{\sc{eant}}}
\newcolumntype{L}[1]{>{\raggedright}p{#1}}
\newcolumntype{C}[1]{>{\centering}p{#1}}
\newcolumntype{R}[1]{>{\raggedleft}p{#1}}

\title{Cosmogenic Backgrounds in Borexino at 3800 m water-equivalent depth}

\collaboration{Borexino collaboration}

\author[a]{G.~Bellini,}
\author[b]{J.~Benziger,}
\author[c]{D.~Bick,}
\author[d]{G.~Bonfini,}
\author[e]{D.~Bravo,}
\author[a]{M.~Buizza Avanzini,}
\author[a]{B.~Caccianiga,}
\author[f]{L.~Cadonati,}
\author[g]{F.~Calaprice,}
\author[d]{P.~Cavalcante,}
\author[g]{A.~Chavarria,}
\author[h]{A.~Chepurnov,}
\author[a]{D.~D{\textquoteright}Angelo,}
\author[i]{S.~Davini,}
\author[j]{A.~Derbin,}
\author[i]{A.~Empl,}
\author[k]{A.~Etenko,}
\author[d,l]{K.~Fomenko,}
\author[m]{D.~Franco,}
\author[g]{C.~Galbiati,}
\author[d]{S.~Gazzana,}
\author[m]{C.~Ghiano,}
\author[a]{M.~Giammarchi,}
\author[o]{M.~G\"{o}ger-Neff,}
\author[g]{A.~Goretti,}
\author[g]{L.~Grandi,}
\author[c]{C.~Hagner,}
\author[i]{E.~Hungerford,}
\author[d]{Aldo~Ianni,}
\author[g]{Andrea~Ianni,}
\author[p]{V.~Kobychev,}
\author[l]{D.~Korablev,}
\author[i]{G.~Korga,}
\author[m]{D.~Kryn,}
\author[d]{M.~Laubenstein,}
\author[o]{T.~Lewke,}
\author[k]{E.~Litvinovich,}
\author[g]{B.~Loer,}
\author[a]{P.~Lombardi,}
\author[d]{F.~Lombardi,}
\author[a]{L.~Ludhova,}
\author[k]{G.~Lukyanchenko,}
\author[k]{I.~Machulin,}
\author[e]{S.~Manecki,}
\author[n]{W. Maneschg,}
\author[q]{G.~Manuzio,}
\author[o]{Q.~Meindl,}
\author[a]{E.~Meroni,}
\author[a]{L.~Miramonti,}
\author[r]{M.~Misiaszek,}
\author[o]{R.~M\"ollenberg,}
\author[g]{P.~Mosteiro,}
\author[j]{V.~Muratova,}
\author[o]{L.~Oberauer,}
\author[m]{M.~Obolensky,}
\author[s]{F.~Ortica,}
\author[f]{K.~Otis,}
\author[q]{M.~Pallavicini,}
\author[e]{L.~Papp,}
\author[q]{L.~Perasso,}
\author[m]{S.~Perasso,}
\author[f]{A.~Pocar,}
\author[a]{G.~Ranucci,}
\author[d]{A.~Razeto,}
\author[a]{A.~Re,}
\author[s]{A.~Romani,}
\author[d]{N.~Rossi,}
\author[g]{R.~Saldanha,}
\author[q]{C.~Salvo,}
\author[o]{S.~Sch\"onert,}
\author[n]{H.~Simgen,}
\author[k]{M.~Skorokhvatov,}
\author[l]{O.~Smirnov,}
\author[l]{A.~Sotnikov,}
\author[k]{S.~Sukhotin,}
\author[t,k]{Y.~Suvorov,}
\author[d]{R.~Tartaglia,}
\author[q]{G.~Testera,}
\author[m]{D.~Vignaud,}
\author[e]{R.B.~Vogelaar,}
\author[o]{F.~von~Feilitzsch,}
\author[o]{J.~Winter,}
\author[r]{M.~Wojcik,}
\author[g]{A.~Wright,}
\author[c]{M.~Wurm,}
\author[g]{J.~Xu,}
\author[l]{O.~Zaimidoroga,}
\author[q]{S.~Zavatarelli,}
\author[r]{G.~Zuzel}

\affiliation[a]{Dipartimento di Fisica, Universit\`{a} degli Studi e INFN, Milano 20133, Italy}
\affiliation[b]{Chemical Engineering Department, Princeton University, Princeton, NJ 08544, USA}
\affiliation[c]{University of Hamburg, Hamburg, Germany}
\affiliation[d]{INFN Laboratori Nazionali del Gran Sasso, Assergi 67010, Italy}
\affiliation[e]{Physics Department, Virginia Polytechnic Institute and State University, Blacksburg, VA 24061, USA}
\affiliation[f]{Physics Department, University of Massachusetts, Amherst 01003, USA}
\affiliation[g]{Physics Department, Princeton University, Princeton, NJ 08544, USA}
\affiliation[h]{Lomonosov Moscow State University Skobeltsyn Institute of Nuclear  Physics, Moscow 119234, Russia}
\affiliation[i]{Department of Physics, University of Houston, Houston,  TX 77204, USA}
\affiliation[j]{St. Petersburg Nuclear Physics Institute, Gatchina 188350, Russia}
\affiliation[k]{NRC Kurchatov Institute, Moscow 123182, Russia}
\affiliation[l]{Joint Institute for Nuclear Research, Dubna 141980, Russia}
\affiliation[m]{APC, Univ. Paris Diderot, CNRS/IN2P3, CEA/Irfu, Obs. de Paris, Sorbonne Paris Cit\'e, France}
\affiliation[n]{Max-Plank-Institut f\"{u}r Kernphysik, Heidelberg 69029, Germany}
\affiliation[o]{Physik Department, Technische Universit\"{a}t M\"{u}nchen, Garching 85747, Germany}
\affiliation[p]{Institute for Nuclear Research, Kiev 03680, Ukraine}
\affiliation[q]{Dipartimento di Fisica, Universit\`{a} e INFN, Genova 16146, Italy}
\affiliation[r]{M. Smoluchowski Institute of Physics, Jagellonian University, Crakow, 30059, Poland}
\affiliation[s]{Dipartimento di Chimica, Universit\`{a} e INFN, Perugia 06123, Italy }
\affiliation[t]{Physics and Astronomy Department, University of California Los Angeles (UCLA), Los Angeles, CA 90095, USA}

\date{\today}

\abstract{
The solar neutrino experiment Borexino, which is located in the Gran Sasso
underground laboratories, is in a unique position to study 
muon-induced backgrounds in an organic liquid scintillator. In this
study, a large sample of cosmic muons is identified and
tracked by a muon veto detector external to the liquid scintillator,
and by the specific light patterns observed when muons cross the
scintillator volume. The yield of muon-induced neutrons is found
to be $Y_{n} = (3.10\pm 0.11) \cdot 10^{-4}\,n/(\mu \cdot
(\rm{g/cm}^{2}))$.   The distance profile between the parent
muon track and the neutron capture point has the 
average value $\lambda=(81.5\pm2.7)$\,cm.   
Additionally the yields of a number of cosmogenic radioisotopes are
measured for $^{12}$N, $^{12}$B, $^{8}$He, $^{9}$C, $^{9}$Li, $^{8}$B, $^{6}$He, $^{8}$Li, $^{11}$Be, $^{10}$C and $^{11}$C. 
All results are compared with Monte Carlo simulation predictions
using the F{\sc{luka}} and \Geant4 packages.
General agreement between data and simulation is observed for the cosmogenic
production yields with a few exceptions, the most prominent case being $^{11}$C yield for which both codes return about $50\%$ lower values. 
The predicted $\mu$-n distance profile and the neutron multiplicity distribution are found to
be overall consistent with data.
}

\keywords{Borexino, Muon, Cosmic, Cosmogenic, Neutron}

\notoc

\begin{document}
\maketitle

\section{Introduction}
\label{sec::intro}

The Borexino experiment is a 278\,t liquid-scintillator detector 
designed for real-time measurements of low energy ($<$20\,MeV) neutrinos. 
The primary goal of the experiment is the spectroscopy of solar neutrinos.
In this respect Borexino has performed a
precision measurement of the {$^7$Be} neutrino line \cite{bx08be7,bx11be7}, has
lowered the threshold for the real-time detection of the {$^8$B} 
neutrino spectrum to 3\,MeV \cite{bx10b8}, and has directly observed
the neutrinos of the pep line, at the same time placing the most
stringent limit on the CNO neutrino flux \cite{bx11pep}.
Borexino is also very competitive in the detection of
anti-neutrinos ($\bar\nu$), having reported a first observation of
geo-neutrinos in 2010 \cite{bx10geo}, followed by a recent new
 measurement \cite{bx13geo}.
Finally, the experiment is sensitive to 
neutrino signals from a galactic core-collapse supernova
\cite{Cadonati:2000kq,Dasgupta:2011wg}. 

The extremely low background in the scintillator target is essential 
to the success of Borexino. While careful pre-selection of detector
materials and extensive purification of the organic scintillator is
necessary, shielding from external, and especially cosmic, radiation is of
comparable importance.  
The detector is located deep underground (3800 meters of
water-equivalent, m w.e.) in the Hall C of the Laboratori Nazionali
del Gran Sasso (LNGS, Italy), where the cosmic muon flux is suppressed
by about six orders of magnitude.  
In spite of this large attenuation factor, residual muons
constitute an important source of background for neutrino detection.  
For example, they produce neutrons or radioactive isotopes by spallation
reactions in target materials, {\it e.g.}~{$^{12}$C}, and these produce 
signals which mimic the observation of a reaction of interest. 

An understanding and mitigation of muon-induced backgrounds are of great
relevance to all investigations of rare processes. In the
majority of underground rare-event experiments, muons and their
spallation products constitute a severe source of background. For
example, in direct dark matter searches, neutron interactions  
feature a signature very similar to those induced by WIMPs, and 
careful shielding must be employed. In $0\nu\beta\beta$ experiments, 
the $\beta$-decays of long-lived radioisotopes produced in-situ can be
significant background components. It is expected that cosmogenic
backgrounds will be even more important in the next generation of
low-background experiments, as the detector sizes and sensitivities are
increasing. Thus more sophisticated muon vetoes as well as
extensive shielding will be necessary.

This paper presents the results of a detailed investigation of muons,
and especially their spallation products, which were detected by
Borexino. Due to its simple geometry, large mass, and excellent event
reconstruction capability, Borexino offers the unique possibility
for a precise study of cosmogenic production inside a large, uniform
volume of low-Z material.  
The paper is structured as follows. After introducing the detector
layout and the general cosmic background conditions found at the LNGS
facility (section~\ref{sec::bx}), we review the results of the cosmic muon flux
and present the reconstructed angular distribution of the muons 
(section~\ref{sec::muons}). We continue with a detailed study of the
neutron production rate and multiplicity in liquid scintillator 
(section~\ref{sec::neutrons}). We also present the lateral distance
profile of neutron captures with respect to the parent muon track.  This is of
special interest for dark matter experiments relying on low-Z
materials for neutron shielding. Moreover, we investigate the
production of a selection of cosmogenic radioisotopes in the
scintillator volume (section~\ref{sec::cosmogenics}).
Finally, we perform a detailed comparison of our experimental
results with the rates and profiles predicted by the commonly used
F{\sc{luka}} and \Geant4 simulation codes (section~\ref{sec::mc}). This
validity check is of considerable importance as Monte Carlo
simulations represent virtually the only means to transfer our
findings to the various detector geometries realized in other
low-background experiments. We also show the 
production yields determined in the KamLAND experiment~\cite{Abe:2009aa}
which features a setup largely comparable to
Borexino. Section \ref{sec::conclusions} summarizes our main results.

\section{The Borexino Detector at the LNGS}
\label{sec::bx}

\begin{figure}
\begin{center}
\includegraphics[width=0.7\textwidth]{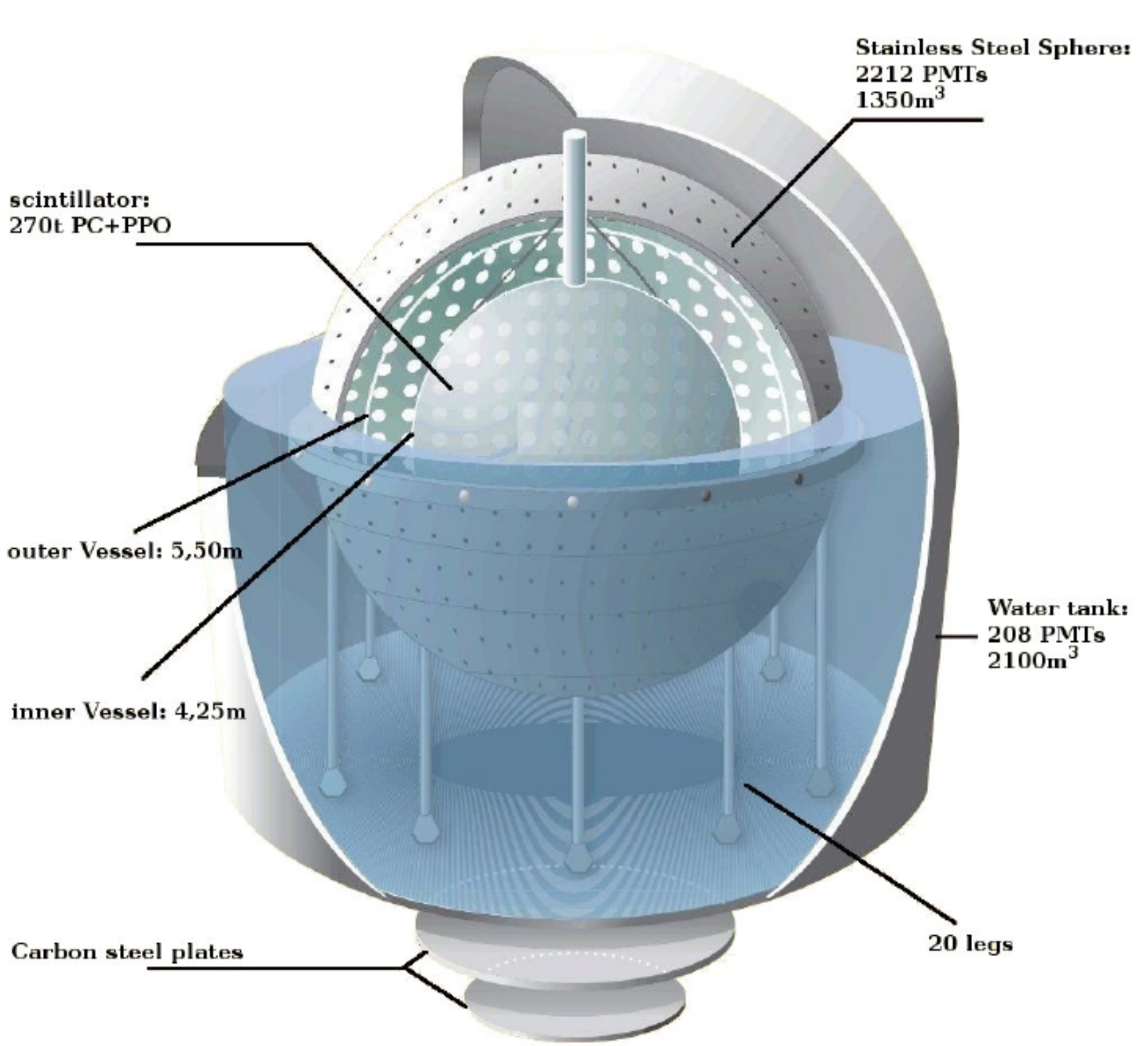}
\caption{Sketch of the Borexino detector.} 
\label{fig:BxDet}
\end{center}
\end{figure} 
 
The Borexino detector \cite{bx08det} consists of a spherical inner
detector (ID) containing a liquid scintillator target and a
surrounding outer detector (OD) consisting of a large water tank. This
tank acts both as
passive shield and as an active muon veto. The general layout is
presented in figure~\ref{fig:BxDet}.  
The central, active scintillator  consists of pseudocumene (PC, 1,2,4
trimethylbenzene), doped with 1.5\,g/liter of PPO (2,5-diphenyloxazole,
a fluorescent dye). The nominal target mass is 278\,t. 
The scintillator is contained in a thin (125\,\textmu m) nylon vessel
of 4.25\,m radius and is shielded by two concentric inactive PC
buffers (323\,t and 567\,t) doped with few g/l of a scintillation light
quencher (dimethylphthalate). 
The two PC buffers are separated by a second thin nylon membrane to
prevent diffusion of radon towards the scintillator. The scintillator
and buffers are contained in a Stainless Steel Sphere (SSS) with a
diameter of 13.7\,m. The SSS is enclosed in a 18.0\,m diameter,
16.9\,m high domed Water Tank (WT), containing 2100\,t of ultra-pure
water. The scintillation light is detected via 2212 8"-photomultiplier
tubes (PMTs) uniformly distributed on the inner surface of the SSS.  
Additional 208 8" PMTs instrument the WT and detect the Cherenkov
light radiated by muons in the water shield \cite{bx11muo}. 
 
In Borexino, low energy neutrinos ($\nu$) of all  flavors are detected
by means of their elastic scattering off electrons or, in case of
electron anti-neutrinos ($\bar\nu_e$), via the inverse $\beta$ decay on
free protons. The electron (positron) recoil energy is converted into
scintillation light which is then collected by the ID PMTs. Borexino
is sensitive to neutrinos of at least 100\,keV in energy, while the
inverse $\beta$ decay induced by anti-neutrinos requires a minimum
neutrino energy of 1.8\,MeV. While cosmic muons crossing the ID deposit
much greater energies and create substantially more light, cosmogenic
neutrons and radioisotopes induce scintillation signals on a scale
similar to neutrino interactions.

\section{Cosmic Muons}
\label{sec::muons}

\begin{figure}
\begin{center}
\includegraphics[width=0.8\textwidth]{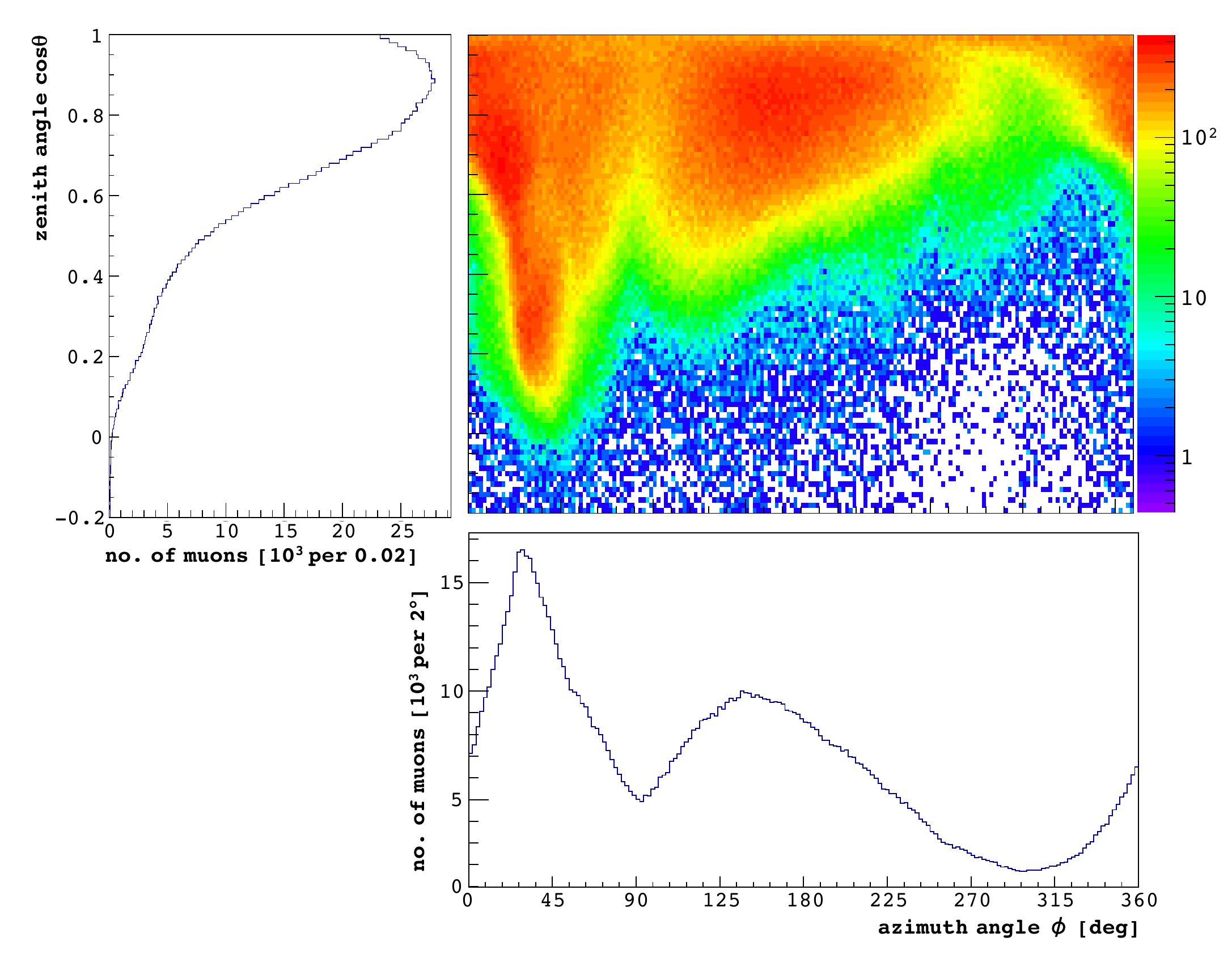}
\caption{Angular distribution of the muons crossing the
  Borexino IV. The side plots show the projections to  astronomic
  azimuth and zenith angles.}  
\label{fig:muonangular}
\end{center}
\end{figure}

The primary cosmic muon flux arriving at Earth's surface
(6.5$\cdot$10$^5$ $\mu$/(m$^2$$\cdot$h)) is strongly attenuated when
penetrating the mountain above the detector by
about a factor 10$^6$. The rock shielding is equivalent to some
3800\,m of water.  
Thus the mean energy of the muons at LNGS site is about 280\,GeV, compared to
about 1\,GeV at the surface, since the lower energy muons incident at the
surface are absorbed, and the spectrum steeply falls as a function
of energy \cite{bib:macro03}.  

We consider in this article only muons generating signals in both the ID and the OD for which 
track reconstruction is also performed. 
Muons identification occurs with three different methods \cite{bx11muo}.  
The first two are based on the detection of the Cherenkov light produced in the water.  
In the first method, the light triggers the OD sub-system (muon trigger flag, MTF). 
In the second method, a cluster of OD PMT pulses is identified, correlated in space and time (muon cluster flag, MCF).
The third method relies on pulse shape
identification of muon tracks among the point-like scintillation
events detected by the ID (inner detector flag, IDF).  
The detection efficiencies are 0.9925(2), 0.9928(2) and 0.9890(1) respectively.
The cosmic muon interaction rate in Borexino was found in
\cite{bx12seas} to be 
($4310\pm2_{\textrm{stat}}\pm10_{\textrm{syst}}$)\,d$^{-1}$ and
corresponds to a flux of $(3.41\pm0.01)\cdot10^{-4}\,{\rm m}^{-2}{\rm s}^{-1}$
as measured in Hall C of the LNGS laboratory. 

The Borexino ID features a uniform
acceptance for incident cosmic muons which is independent of the arrival
directions, thanks to its spherical symmetry. The observed angular
distribution is shown in 
figure~\ref{fig:muonangular} as a function of the astronomic azimuth
($\phi$) and zenith ($\theta$) angles in a two-dimensional contour
plot\footnote{A table listing the numerical values of the distribution has been included in the supplementary materials.}. 
In addition, the one-dimensional projections on the $\theta$ and
$\phi$-planes are also presented. All three distributions reflect the
influence of the local mountain topology: The differences in the
thickness of the overlaying rock  are imprinted as
angle-dependent variations in the residual muon flux. Note that due to
its uniform detection efficiency, Borexino is in a unique position to
map the muon distribution at depth close to the horizon without angular
distortions. 

To obtain these plots, the ID tracking algorithm described in
\cite{bx11muo} was employed.  These plots consider only muons crossing the
IV because their tracks offer the best angular resolution. For this
selection, the (redundant) requirements of a reconstructed energy
deposition of at least 300\,MeV in the IV and an impact parameter
of less than 4.25\,m were applied. The remaining sample consists of
1\,221\,470 individual muon tracks.

\section{Cosmogenic Neutrons}
\label{sec::neutrons}

Cosmic muons crossing the detector produce fast neutrons through
different spallation processes on carbon nuclei.
The neutron velocities are slowed in the scintillator by collisions
with hydrogen or carbon nuclei to thermal sub-eV energies in a few
scatterings. This process occurs within a few tens of ns. 
Consequently, signals from ionizations due to the recoiling nuclei 
cannot be disentangled from the much higher light emission of an incident
muon.  For the same reason, fast neutron captures 
are also not visible. 
About 1\% of the neutrons are
captured during a fast capture process, as determined by our
Monte Carlo simulations. 
 We do not correct our results to account for this
effect in this paper, and hereafter we refer to all neutron captures
as due to slower thermal capture. 
The mean capture time of a thermal neutron in the liquid scintillator
is $\sim$250\,\textmu s, and the subsequent gamma-ray emission is
distinctly visible.
The energy emitted in gamma rays is 2.2\,MeV if the neutron is captured on
hydrogen and 4.9\,MeV if captured on carbon.  
Based on the elemental composition of the scintillator and the
relative capture cross-sections,  
about 99\% of all thermal neutron captures are expected on hydrogen
\cite{IAEA}. 
The ability of Borexino to detect cosmogenic neutrons was
described in detail in \cite{bx11muo}. 

\subsection{Neutron detection with the main electronics}
\label{sec:neutrons_mainelec}

The data acquisition system
issues a dedicated neutron acquisition gate of 1.6\,ms length after a
muon crosses the SSS. The captured PMT
pulses are due to neutron capture-gammas and accidental
$^{14}$C decays.  The latter is an intrinsic contaminant of the
scintillator.  Particular care must be taken when the muon crosses the
scintillator because of the large amount of visible light which is
created. For these events, the baseline of the front-end electronics
takes up to 30\,\textmu s to stabilize. 
 Furthermore as a result of the
intense PMT illumination, the front of the acquisition gate is
highly populated by noise pulses creating a variable baseline on
which the other pulses are superimposed. Thus the
active time window for the analysis is set after the
baseline stabilizes (30\,\textmu s). An ad-hoc
algorithm detects physical events by identifying \emph{clusters} of
time-correlated PMT pulses on top of the baseline with high
efficiency. An energy threshold of 1.3\,MeV is imposed via a
variable selection cut based on detector saturation.
The overall detection efficiency above threshold
was determined using neutron
samples which were selected as described below. The result
is $\varepsilon_{\textrm{det}}= (91.7 \pm 1.7_\textrm{stat} \pm
0.9_\textrm{syst})\,\%$ for cosmogenic neutrons captured later than
30\,\textmu s after their parent muon is observed.  The error 
incorporates the uncertainty in the threshold definition of the energy. 

The data sample used for the analysis contains 559 live days, covering
the time period from January 6, 2008, until February 2, 2010. The
respective 
average scintillator volume contained in the nylon vessel was
measured to be $(306.9\pm 2.9)\,\textrm{m}^3$, or $(22.8 \pm 0.2)\,\%$
of the SSS volume. This translates into an active mass of $(270.2 \pm
2.6)$\,t. The volume in which neutron captures are detected is not
identical with that defined by the physical vessel 
boundaries.  For example, gamma-rays generated in the buffer close to the nylon
vessel may cross into the scintillator volume and create a sizeable
light output, while gamma-rays from neutron captures inside the
scintillator may escape into the buffer 
undetected. This effect was quantified using a Monte Carlo
simulation of neutron capture gammas in both buffer and scintillator.  The
simulation included the transport of scintillation light and electronic 
effects. Of all $\gamma$'s
generated inside the SSS, the fraction of 
events which deposit more than 1.3\,MeV energy in the IV,
$\varepsilon_{\rm vol}$, is $(23.0 \pm 0.3)\,\%$  for captures on
hydrogen, and $(24.2 \pm 0.2)\,\%$ for captures on carbon. Both the
physical volume evaluation and the Monte Carlo 
simulation take into account variations in actual vessel shape and size over the
data collection period. 

The neutron capture time and the purity of the sample were
studied based on the time difference ($\Delta t$) between the
occurrence of candidate clusters and their parent muons. The $\Delta
t$ distribution is fit by the sum of an exponential decay and a
flat component for uncorrelated events. The resulting neutron capture
time is $\tau_n = (259.7 \pm 1.3 _\textrm{stat} \pm 2.0
_\textrm{syst})$\,\textmu s, and is in agreement with our previous
measurement \cite{bx11muo}. Based on this value, the fraction of
neutrons captured later than 30\,\textmu s after the parent muon is
$\varepsilon_{\rm t}=(89.1\pm0.8)\,\%$. The contamination by
uncorrelated events is found to be $(0.5\pm0.2)\,\%$. 

\subsection{Neutron detection with waveform digitizers}
\label{sec:neutrons_fwfd}

The Borexino detector is equipped with auxiliary DAQ systems based on
fast waveform digitizers. 
Two of these systems (hereafter SYS1 and SYS2) were used in this
analysis to evaluate the neutron detection efficiency
$\varepsilon_{\textrm{det}}$ of the main data acquisition system.
Furthermore, these two systems 
provide a cross check on the neutron yield measurement
(section~\ref{sec:neutrons_rates}). 

SYS1 is a single channel 500\,MHz, 8\,bit digitizer (Acqiris DP235)
which records the cumulative analog output
of all ID PMTs. It is triggered by the MTF 
condition of the outer detector, and collects data for about
1.6\,ms. A cluster-finding algorithm identifies gamma-ray capture
signals between 30 and 1590\,\textmu s after muon detection.  
An energy threshold of 1.3\,MeV is applied to reject noise pulses. 
SYS1 collected data between April 2008 and November 2009. 
The fit to the $\Delta t$ distribution returns a capture time of
$(261\pm1_\textrm{stat}\pm1_\textrm{syst}$)\,\textmu s.  
An additional flat component which would be allowed by the fit is 
consistent with zero at the 1$\sigma$ confidence level.  

SYS2 is an auxiliary hardware architecture, based upon 96 waveform
digitizers (CAEN v896: 400\,MHz, 8\,bit). 
Each channel receives the analog sum of 16 or 24 PMTs. 
The system operates independently from the main DAQ.
A separate trigger is implemented by an FPGA unit (CAEN v1495).
Data used for this analysis were collected between December 2009 and March 2012.
A neutron detection efficiency of essentially 100\,\% is reached for
an energy threshold above 1\,MeV. 
For SYS2, the fit to the $\Delta t$ distribution returns a capture
time of $(258.7 \pm 0.8 _\textrm{stat} \pm 2.0
_\textrm{syst})$\,\textmu s.  
A residual of uncorrelated background on the level of $(0.5\pm0.1)\,\%$
was determined by a fit and when using a delayed time window. 
 
The values of the capture time obtained by SYS1 and SYS2 are in 
good agreement with the values using the main electronics (section~\ref{sec:neutrons_mainelec}).

\subsection{Neutron production rate and multiplicity}
\label{sec:neutrons_rates}

The data set acquired by the main DAQ contains a sample of
$N_{\mu}=2\,384\,738$ muons and $N_{n}=111\,145$ neutrons. 
The neutron capture rate for the efficiencies described above,
is determined to $R_{n}= (90.2 \pm
2.0_\textrm{stat} \pm 2.4_\textrm{syst})\,$(d\;100t)$^{-1}$ after
scaling. The systematic
uncertainty of this measurement is dominated by the neutron capture
time and the average scintillator volume contained inside the IV. 

The rate of muons which produce neutrons that eventually are captured 
inside the IV, is $R^{\mu}_{n}=(67.5 \pm 0.4_\textrm{stat} \pm
0.2_\textrm{syst})\,$d$^{-1}$ ($\sim$1.5\,\% of muons crossing the
ID). 
The detector-specific ratio $R_{n}/R^{\mu}_{n}$ corresponds to an
average neutron multiplicity within the IV volume of
$\overline{M}=(3.61\pm 0.08_\textrm{stat}\pm
0.07_\textrm{syst})\,n/\mu$. 
The distribution of the multiplicities of detected neutron captures is
shown in figure~\ref{fig:multiplicity} where it is compared to Monte
Carlo predictions. 
The neutron yield per unit length of muon track in the target medium is
\begin{eqnarray}\label{eq::neutron_yield}
Y_{n} &  = & \frac{N_{n}}{N_{\mu}}\cdot
\frac{1}{\ell^{\textrm{avg}}_{\mu}}\cdot\frac{1}{\rho_{\textrm{scint}}}
\cdot \frac{1}{\varepsilon_{\textrm{det}}\cdot\varepsilon_{\textrm{t}}
  \cdot \varepsilon_{\textrm{vol}}} \nonumber\\ 
& = & (3.10 \pm 0.07_\textrm{stat} \pm 0.08_\textrm{syst}) \cdot
10^{-4}\,n/(\mu \cdot (\rm{g/cm}^{2})) 
\end{eqnarray}
where $\ell^{\textrm{avg}}_{\mu}=4/3\,R_{\textrm{SSS}}$ is the average
muon path through the SSS with $R_{\textrm{SSS}}=(6.821\pm0.005)$\,m
and $\rho_{\textrm{scint}}=0.88$\,g/cm$^3$ is the scintillator
density. We chose to consider the muon path through the SSS and
not just the IV and to include the ratio of the two volumes in
$\varepsilon_{\textrm{vol}}$ in order to correctly account for the
effective neutron detection volume.  This is discussed in
section~\ref{sec:neutrons_mainelec}. The statistical uncertainty
associated with this result has been assessed by a toy Monte Carlo
code which simulates neutron production by muons in order to observe the size of
the fluctuations.  
For varying values of the muon rate, the statistical uncertainty is
approximately seven times the square root of the number of neutron
captures divided by the live time.  This results in a statistical
uncertainty of 3\%. 

As a consistency check, the neutron yield has been also determined
based on the waveform digitizers: $Y_n^{\rm SYS1}=3.19 \pm
0.08_\textrm{stat} {}^{+0.10}_{-0.09} {}_\textrm{syst}$ for SYS1 and
$Y_n^{\rm SYS2}=2.87 \pm 0.07_\textrm{stat} \pm 0.15_\textrm{syst}$
for SYS2 in units of  $10^{-4}\,n/(\mu \cdot (\rm{g/cm}^{2}))$.  This is in
reasonable agreement with the value given in equation (\ref{eq::neutron_yield}). 

\subsection{Neutron lateral distance} 
\label{sec::LDneutrons}

\begin{figure}
\begin{minipage}{0.5\textwidth}
\includegraphics[width=\textwidth]{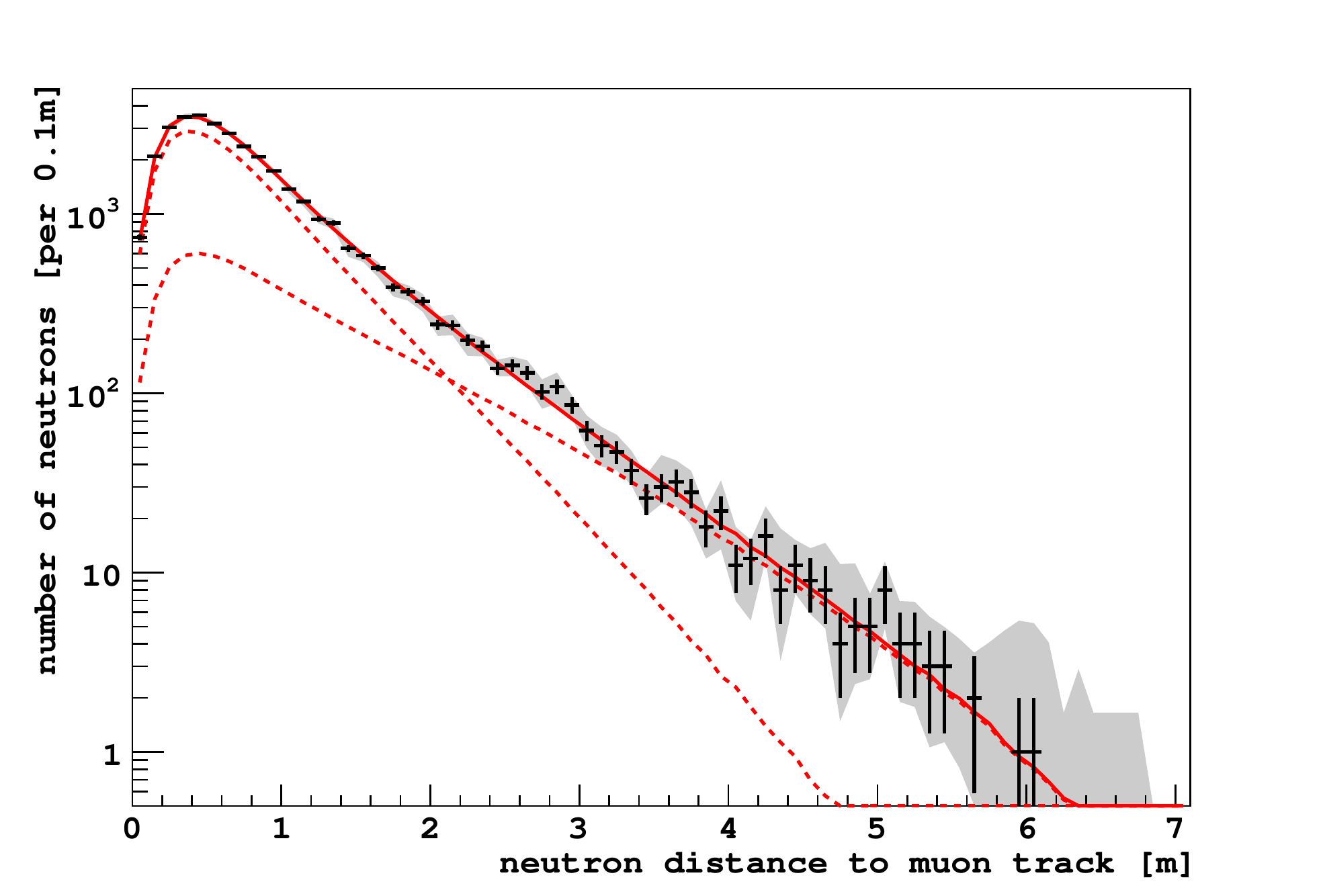}
\end{minipage}
\begin{minipage}{0.5\textwidth}
\centering
\begin{tabular}{rc}

parameter & fit value  \\
\hline
 $\sigma_{\rm muon}$ [cm] & 33.6$\pm$0.6$\pm$0.6 \\
 $\mu_{\rm muon}$ [cm] & 23.8$\pm$0.6$\pm$0.6 \\
 $\lambda_{\rm short}$ [cm] & 61.2$\pm$0.6$\pm$2.6 \\
 $f_{\rm short}$ [\%] & 76.5$\pm$0.5$\pm$5.5 \\
  $\lambda_{\rm long}$ [cm] & 147$\pm$3$\pm$12 \\
 $f_{\rm long}$ [\%] & 23.5$\pm$0.4$\pm$5.5\
\end{tabular}
\end{minipage}
\caption{The muon-neutron distance distribution observed in Borexino:
  black crosses represent the data points for the standard neutron hit
  multiplicity cut.  The shaded-grey area indicates the systematic
  uncertainty. The fit of the toy Monte Carlo is indicated by the solid red
  line. The dashed lines correspond to two exponential components,
  each featuring a decay length $\lambda$ and a relative fraction
  $f$. The muon resolution parameters $\mu$ and $\sigma$ are
  left free in the fit procedure (see sect.~\ref{sec::LDneutrons} and
  \cite{bx11muo} for details). The table lists the best-fit results
  with  statistical and systematic uncertainties. The fit returns
  $\chi^2/{\rm ndf}=57/54$.}  
\label{fig:munedist}
\end{figure}

Borexino can spatially reconstruct the emission point of the
neutron capture gamma-ray as well as the track of its parent
muon. This information may be used to compute the shortest
(perpendicular) distance between the neutron vertex and the parent
muon track (hereafter: neutron lateral distance). A toy
Monte Carlo was fit to the data in order to separate the distance
travelled by the neutrons from resolution effects of the detector (see
below).  

Muon track and neutron verticies were reconstructed based on the Borexino main
DAQ in order to obtain the lateral distance distribution from the data.
Only muons and neutrons having a radial distance less than 4\,m from the
detector center were selected in order to ensure a well-defined geometry for 
comparison to the Monte Carlo.
Moreover, the samples were cleaned to remove tracks and vertices
of inferior reconstruction quality. In case of the muons, only events
which feature spatially compatible tracks in both sub-detectors are
used.  
The neutron selection cuts are much more
restrictive.  Due to electronic effects and PMT afterpulses which
were present after very luminous muons, the spatial reconstruction of
subsequent neutron events can be severely compromised.  This 
results in systematic shifts increasing or decreasing the distance between the
neutron capture point and the parent muon track 
~\cite{bx11muo}. Systematic studies demonstrated  that the majority of
these effects (e.g.~afterpulses)  subside in the first
200\,\textmu s after a muon event. At later times, 
a fraction of the electronic channels might be affected by
buffer overflow, which leads to asymmetric PMT hit patterns. Therefore, 
only neutron captures with a minimum time delay of
200\,\textmu s are compared to a parent muon with the further restriction
that the neutron event was composed of at  least
100 individual PMT signals in order to ensure an unbiased vertex reconstruction
($N_{\rm hits}>100$). The latter condition is easily met by neutrons
produced by minimum-ionizing muons, but depletes the sample of useful
neutron captures in case of very luminous muon events. Such muons are
expected to create extensive hadronic showers, and there is a
risk that the hit multiplicity cut for the neutrons
introduces a bias to the selected sample which preferentially suppresses
neutrons at large distances from their parent tracks.  
Finally, we limit the visible energy window to
$E_\mathrm{vis}\in[0.9;4.8]$\,MeV in order to select only neutron
captures on hydrogen and carbon, while removing a minor contamination
from short-lived cosmogenic isotopes. The combination of cuts reduces
the remaining sample to $\sim$20\,\% of the original neutrons.

The resulting lateral distance distribution is shown in
figure~\ref{fig:munedist}. 
The grey shaded area corresponds to the systematic uncertainty
introduced by the cut $N_{\rm hits}>100$, and was obtained by varying
the minimum $N_{\rm hits}$ condition for neutron selection from 0 to
200.  
Due to the broad initial energy spectrum of the spallation neutrons and the
corresponding distribution of the neutron mean free paths, a simple
exponential law proves insufficient to reproduce the distribution. We find that
at least two exponential components
($\lambda_\textrm{short}$ and $\lambda_\textrm{long})$ are required for
a satisfactory description of the data. The fit function shown in figure~\ref{fig:munedist} was
obtained by a toy Monte Carlo simulation. Apart from the 
exponential components, the fit takes into account the muon and
neutron spatial resolutions, which includes the average displacement 
of the neutrons during thermalization and the finite propagation
distance of the capture gamma in scintillator
($\sim$20\,cm). The geometric impact of the applied radial cuts 
described above are included. The muon lateral
resolution is described by a Gaussian smearing $\sigma$ with a constant
radial offset $\mu$. These are free parameters in the
fit. Conversely, the neutron vertex resolution is set to a fixed value
of 23\,cm (see \cite{bx11muo} for details). 

The fit returns a short component
$\lambda_\textrm{short}=(61.2\pm0.6_{\rm stat}\pm2.6_{\rm syst})$\,cm which is in
agreement with earlier LVD results \cite{lvd99}. The long component
is found to be
$\lambda_\textrm{long}=(147\pm3_{\rm stat}\pm12_{\rm syst})$\,cm.  
Systematic uncertainties for the parameters were determined by
multiple repetitions of the fit while varying the minimum $N_{\rm
  hits}$ condition for the neutrons. Based on the relative weights of
the two effective components, an 
average lateral distance of $\lambda = (81.5\pm2.7)$\,cm was determined. 

\section{Cosmogenic Radioisotopes}
\label{sec::cosmogenics}

\begin{table}
\begin{center}
{\footnotesize \begin{tabular}{|l|c|c|c||l|c|c|c|}
\hline
\bf{Cosmogenic} 				& \bf{Lifetime}   &
\bf{Q-Value}			& \bf{Decay} 			&
\bf{Cosmogenic} & \bf{Lifetime}     & \bf{Q-Value}		&
\bf{Decay}\\ 
\bf{Isotope}						&
& [MeV]       			& \bf{Type}   		&
\bf{Isotope}						&
&  [MeV]       	 & \bf{Type}\\ 
\hline
{\bf $^{12}$N}          & 15.9\,ms			  &  17.3
&  	$\beta ^-$    & {\bf $^{6}$He}          & 1.16\,s
&  3.51 &  					$\beta ^-$ \\ 
{\bf $^{12}$B}          & 29.1\,ms        &  13.4
&  	$\beta ^+$    & {\bf $^{8}$Li}          & 1.21\,s
&  16.0 &  					$\beta ^-$ \\ 
{\bf $^{8}$He}          & 171.7\,ms       &  10.7
&  	$\beta ^-$    & {\bf $^{11}$Be}         & 19.9\,s
&  11.5 &  					$\beta ^-$ \\ 
{\bf $^{9}$C}           & 182.5\,ms       &  16.5
&  	$\beta ^+$    & {\bf $^{10}$C}          & 27.8\,s
&  3.65 &  					$\beta ^+$ \\ 
{\bf $^{9}$Li}          & 257.2\,ms       &  13.6
&  	$\beta ^-$    & {\bf $^{11}$C}          & 29.4\,min
&  1.98 &  					$\beta ^+$ \\ 
{\bf $^{8}$B}           & 1.11\,s    			&  18.0
&  	$\beta ^+$    &
&
&				&
\\ 
\hline
\end{tabular}}
\end{center}
\caption{List of cosmogenic isotopes expected to be produced by muons
  in organic scintillators in measurable rates.} 
\label{tab:list_of_cosmogenics}
\end{table}

In addition to neutrons, radioactive isotopes are produced in
muon-induced spallation processes on the target nuclei. 
A list of the relevant cosmogenic isotopes with their 
properties and sorted by increasing lifetime can be found in table~\ref{tab:list_of_cosmogenics}. 

Candidate events are selected via two observables: the visible energy, $E$, and the time difference, $\Delta t$, with respect to a parent muon. 
The distributions of the two observables are fit simultaneously in an unbinned likelihood fit.
However, matching to a parent muon is not unique in general, as many
muons can be present within the selected analysis time gate $t_g$.  This
results in multiple values of $\Delta t$ for a given candidate, with only one
muon physically correlated  with the neutron decay. This effect is most 
prominent for the analyses on cosmogenic isotopes with lifetimes on the 
order of seconds or longer since in  average cosmic muons
cross the ID every 20\,s. The distribution in $\Delta t$ is fit with the
function: 
\begin{equation}	\label{formula::time_profile}
F(\Delta t) =  \sum\limits_{i}{}\frac{N_i^t}{\tau_i} e^{-\frac{\Delta
    t}{\tau_i}} + \frac{N_{b}^t}{t_g} + \frac{N_{um}^t}{t_g} 
\end{equation}
The number of decays in each isotope profile is $N_i^t$ with
${\tau_i}$ its lifetime.  
Flat contributions, $N_{b}^t$ and $N_{um}^t$, account
for uncorrelated background events and for physically uncorrelated
matches, respectively. The latter is a property of the selected data
set and calculated independently.  
The fit function is valid for time scales much
shorter than the average run duration of a data set ($\sim$6\,h). 
This is valid for all cosmogenic isotopes with the exception of 
$^{11}$C (table~\ref{tab:list_of_cosmogenics}).
 As will be presented in section~\ref{subsec::11C}, a distortion
of the time profile due to run-boundary effects can be avoided by a time
cut of candidate events which occur close to run-start. 
 The spectral shapes of the respective isotopes are generated with the
 \Geant4 based Borexino Monte Carlo code. The simulation reproduces
 the full detector response, yielding about 500\,photoelectrons/MeV of deposited
 energy in $\beta$ and $\gamma$ decays, and spectral fits can be performed 
directly on the number distribution of photoelectrons from
 candidate events.   For easier reference, this
 conversion factor will be used to refer to the energy selection cuts in the
 [MeV] units in the following analyses. Based on
 the visible energy $E$ of an event, the spectral fit function $G(E)$
 is given by\,: 
\begin{equation}	\label{formula::energy_distribution}
G(E) =  \sum\limits_{i}{}{N_i^E} g_i(E) + N_{b}^E g_{b}(E)
\end{equation}
The spectral shapes of the analyzed isotopes are denoted by $g_i(E)$
with the respective number of decays $N_i^E$.  The uncorrelated 
background is addressed with the spectral shape $g_{b}(E)$ and
$N_{b}^E$ is the number of entries. The energy distribution is
generated from events which occur within a time interval, $t_E$, relative
to preceding muons. To enhance the signal to noise ratio, $t_E$ is
chosen to be in the order of the lifetimes of the respective
cosmogenic radionuclides. The number of isotope decays in the time and
energy fit ($N_i^t$ and $N_i^E$), and the background events
($N_b^t$ and $N_b^E$), are related to the known selection cut
efficiencies in time and energy.
 
Most cosmogenic isotopes are expected to be produced at a very low rate.
To reduce accidental coincidences, muons are removed from the sample of candidate isotope events by the application of the MTF and IDF muon identification methods (section \ref{sec::muons}).
The efficiency and small distortions due to the application of IDF are included in the Monte Carlo generated spectral shapes. 
MTF efficiency is accounted for \emph{a posteriori}.
Unless stated otherwise, the positions of candidate events 
are also required to lie within a \textit{fiducial volume} (FV). 
This is defined by a sphere of a $3$\,m radius which corresponds to a 99.6\,t mass of liquid scintillator. 
The systematic uncertainty in the reconstructed volume for the decay energies of
interest is estimated to $\pm$\,3.8\,\% and contributes to the uncertainty of all measured yields.
Except where differently noted, 
the analyses on cosmogenic radioisotopes are based on the same data set used for the
neutron yield analysis (section \ref{sec:neutrons_mainelec}).

\subsection{$^{12}$N and $^{12}$B}
\label{subsec::12N_and_12B}

\begin{figure}
\begin{center}
\includegraphics[width=0.7\textwidth]{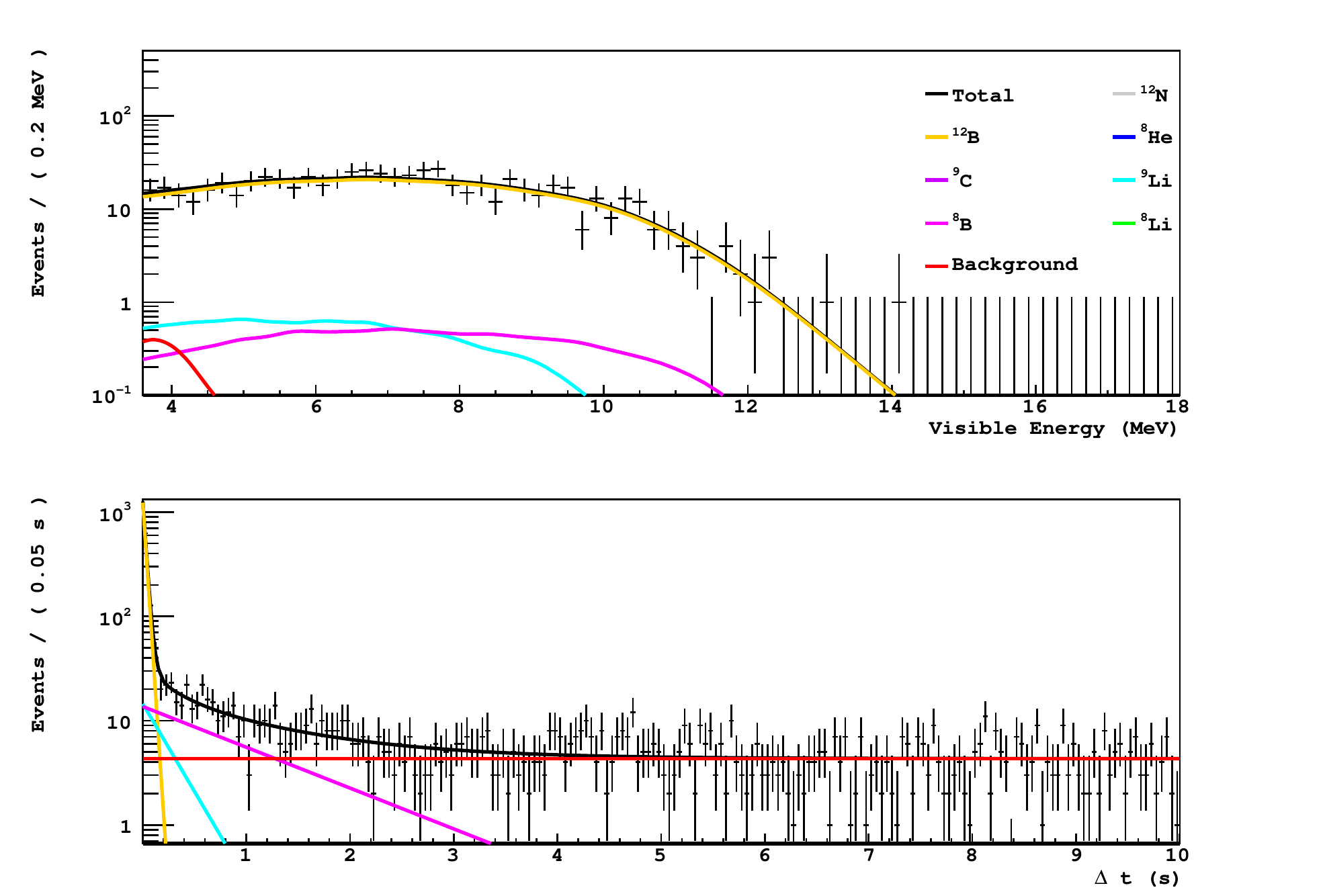}
\caption{Simultaneous fit of the cosmogenic isotopes $^{12}$N
  and $^{12}$B in visible energy deposition (top panel) and decay
  time relative to preceding muons (bottom panel), including the isotopes
  $^{8}$He (blue), $^{9}$C, $^{9}$Li, $^{8}$B
  and $^{8}$Li as contaminants. The fit returns only upper
  limits for the isotopes $^{12}$N, $^{8}$He, $^{9}$C and $^{8}$Li,
  and they cannot be seen in the graph.
The goodness of the simultaneous fit is $\chi^2/\textrm{ndf} = 348/236$.}
\label{fig:12N_and_12B}
\end{center}
\end{figure}

\begin{figure}
\begin{center}
\includegraphics[width=0.7\textwidth]{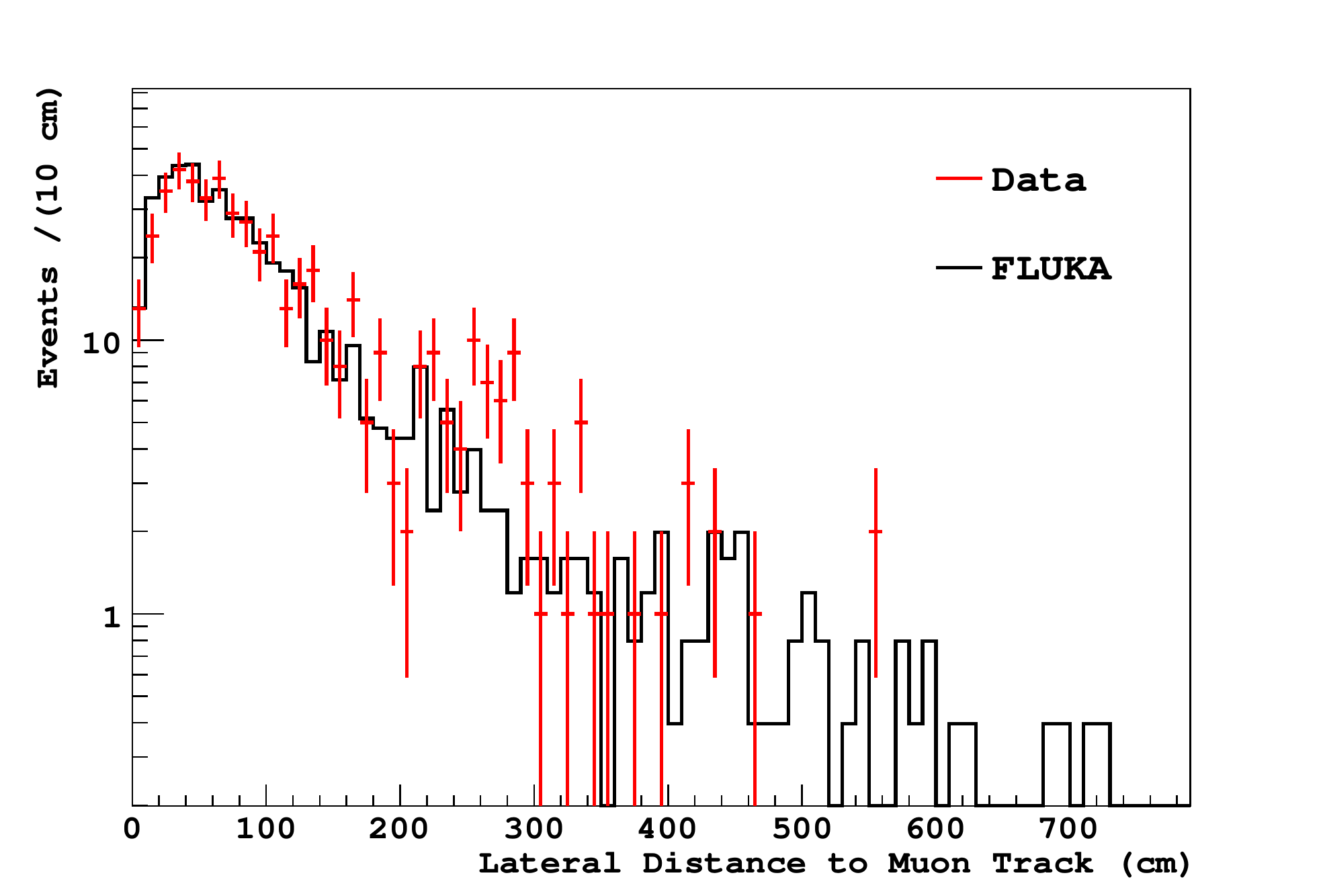}
\caption{Comparison of measured and simulated  lateral production profiles for
 cosmogenic $^{12}$B candidates inside the FV with respect to the parent muon
 track.  For improved spatial resolution we select only tracks of muons which
 cross the IV at an impact parameter of less than 4\,m.} 
\label{fig:12B_distance}
\end{center}
\end{figure}

Candidate events for the decays of $^{12}$N ($\beta ^+$-emitter, $\tau
 = 15.9$\,ms, $Q = 17.3$\,MeV) and $^{12}$B ($\beta ^-$-emitter, $\tau
 = 29.1$\,ms, $Q = 13.4$\,MeV) are selected within an energy range
 $E\in[3.6,18]$\,MeV and a time gate $\Delta t \in t_g =
 [2$\,ms$,10$\,s$]$ with respect to a preceding muon event. 
 The energy distribution is constructed from events with $\Delta t \in t_E =
 [2,60]$\,ms.  This increases the signal-to-background ratio. After
 each muon, a $2$\,ms veto is applied to avoid muon-induced
 secondaries (mainly neutrons) which leads to a negligible dead
 time. 
Furthermore, decays of the cosmogenic isotopes $^{8}$He, $^{9}$C,
 $^{9}$Li, $^{8}$B and $^{8}$Li are considered as contaminations and
are fit alongside $^{12}$N and $^{12}$B. The upper limit of the time
 gate $t_g$ (i.e. 10\,s) is driven by the lifetimes of the isotopes
 $^{8}$B ($\tau = 1.11$\,s) and $^{8}$Li ($\tau = 1.21$\,s).
In addition, a fraction of $(86.2 \pm 0.2)\,\%$ of all $^{11}$Be decays
 is expected within the selected energy range. However, due to its low
 production rate and long lifetime ($\tau = 19.9$\,s), this
 contribution is estimated to be less than 
1\% in the time profile, and negligible in the energy distribution
 (section~\ref{subsec::11Be}). The background spectral shape is built
 from events with $\Delta t > 10$\,s to avoid accidental coincidences of
 short-lived cosmogenic isotopes. 
Figure~\ref{fig:12N_and_12B} shows the simultaneous fit in energy and
 time. The efficiency of the energy cut is evaluated via the Borexino
 Monte Carlo simulation to be $\varepsilon(^{12}$N$) = (79.3 \pm 0.4)\,\%$
 and $\varepsilon(^{12}$B$) = (84.0 \pm 
0.3)\,\%$.  The uncertainties are due to the detector energy
 resolution. The simultaneous fit yields a rate of $R(^{12}$N$) <
 0.03\,$(d\;100t)$^{-1}$ at a 3$\sigma$ confidence level, and
 $R(^{12}$B$) = (1.62 \pm 0.07_\textrm{stat} \pm
 0.06_\textrm{syst})\,$(d\;100t)$^{-1}$. These translate to production
 yields of $Y(^{12}$N$) < 1.1 \cdot 10^{-7}\,/(\mu \cdot
 (\rm{g/cm}^{2}))$ and $Y(^{12}$B$) = (55.6 \pm 2.5_\textrm{stat} \pm
 2.1_\textrm{syst})\cdot 10^{-7}\,/(\mu \cdot (\rm{g/cm}^{2}))$.

$^{12}$B at a level of $(94.6 \pm
 0.3)\,\%$ of all events is clearly the dominating cosmogenic isotope 
within the selected time window ($\Delta t \in [2,60]$\,ms). 
Figure~\ref{fig:12B_distance} shows the lateral
 production profile of this isotope with respect to the reconstructed
 track of the parent muon.

\subsection{$^{8}$He and $^{9}$Li}
\label{subsec::8He_and_9Li}

\begin{figure}
\begin{center}
\includegraphics[width=0.7\textwidth]{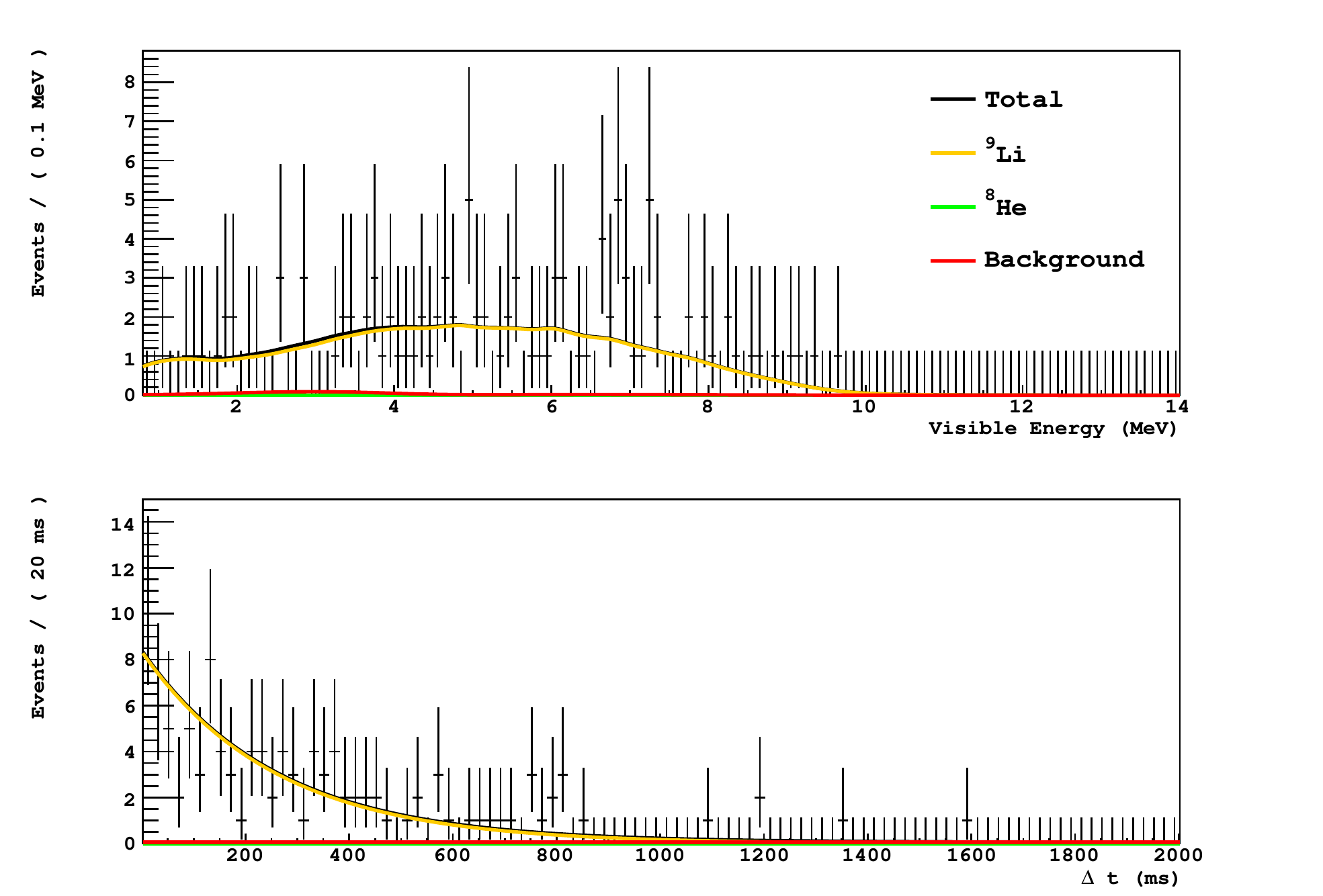}
\caption{Simultaneous fit of the cosmogenic isotopes $^{8}$He 
  and $^{9}$Li in visible energy deposition (top panel) and decay
  time relative to preceding muons (bottom panel). The fit returns only an
  upper limit for the isotope $^{8}$He
  and its line cannot be seen in the graph.
  The goodness of the simultaneous fit is $\chi^2/\textrm{ndf} =71/98$.} 
\label{fig:8He_and_9Li}
\end{center}
\end{figure}

Both $\beta^-$-emitters $^{8}$He ($\tau = 171.7$\,ms , $Q =
10.7$\,MeV) and $^{9}$Li ($\tau = 257.2$\,ms, $Q = 13.6$\,MeV) exhibit
daughter nuclei with neutron-unstable excited states. With a 16\%
branching ratio, the $\beta$-decay of 
$^{8}$He populates such a state in $^{8}$Li. For $^{9}$Li, the
branching ratio to a neutron-unstable state in $^{9}$Be is 51\%. The
subsequently emitted neutron is captured mainly on hydrogen with a
mean capture time of $(259.7 \pm 
3.3)$\,\textmu s (section~\ref{sec:neutrons_mainelec}), emitting a
characteristic $2.2$\,MeV gamma-ray. The triple-coincidence of a muon,
a $\beta$-emission, and a delayed neutron capture provides a very clean
signature, which allows analysis of events within 
the entire mass of the inner vessel. Thus, an enlarged data set of 1366 live
days taken between January 6, 2008, and August 31, 2012, is used for
the analysis with a mean inner vessel mass of $(268.2 \pm
2.8)$\,t.  
The requirements for candidate events for the $\beta$-emissions are
$\Delta t \in [2\,$ms$,2\,$s] and $E\in[0.8,14]$\,MeV. The energy
distribution is taken from events with $\Delta t \in
[$2\,ms,1\,s]. Subsequent neutron capture 
candidates are selected by $E\in[1.7,2.6]$\;MeV. These are required to occur
within $1$\,m distance and a maximum time delay of $1.3$\,ms to
a $\beta$-like event. The uncorrelated background spectrum is
derived from events with $\Delta t > 1$\,s.  
Figure~\ref{fig:8He_and_9Li} shows the simultaneous fit in energy and
time of $^{8}$He and $^{9}$Li. The $\beta n$ selection cut efficiency
has been evaluated to be $\varepsilon(\beta n) = (79.3 \pm 0.4)\,\%$ via
the Borexino Monte Carlo simulation. The energy cut efficiencies are
estimated to $\varepsilon(^{8}$He$) = (99.49 \pm 0.05)\,\%$ and
$\varepsilon(^{9}$Li$) = (96.99 \pm 0.11)\,\%$.  
The fit returns only an upper limit for $^{8}$He.
Isotope production rates of $R(^{8}$He$) < 0.042\,$(d\,100t)$^{-1}$ 
 at a $3\sigma$ confidence level, and $R(^{9}$Li$) = (0.083 \pm
0.009_\textrm{stat} \pm 0.001_\textrm{syst})\,$(d\,100t)$^{-1}$ are observed. The
corresponding yields are $Y(^{8}$He$) < 1.5 \cdot 10^{-7}\,/(\mu \cdot
(\rm{g/cm}^{2}))$ and $Y(^{9}$Li$) = (2.9 \pm 0.3)\cdot 10^{-7}\,/(\mu
\cdot (\rm{g/cm}^{2}))$.

\subsection{$^{8}$B, $^{6}$He and $^{8}$Li}
\label{subsec::8B_6He_and_8Li}

\begin{table}
\begin{center}
{\footnotesize
\begin{tabular}{|c|c|c|c|c|c|}
\hline
\textbf{Energy regime} & \textbf{Time Gate $t_g$} & \textbf{Time Gate $t_E$} &  \textbf{Cosmogenic} & \textbf{Lifetime} & \textbf{Energy Cut } \\
                                    &   [s]                                  &   [s]                                   & \textbf{Isotope}         & [s]                & \textbf{Efficiency} $[\%]$ \\
\hline
                                    &                                        &                                                  & $^{8}$B                         & 1.11                       & $4.0 \pm 0.3$ \\
ER1                                 & $[1,140]$                              & $[1,2]$                          & $^{6}$He                & 1.16                       & $16.8 \pm 0.3$ \\
$E \in [2,3.2]$\,MeV        &                                        &                                          & $^{8}$Li                        & 1.21                         & $8.1 \pm 0.2$ \\
                                    &                                        &                                            & $^{10}$C              & 27.8                        & $77.1 \pm 0.2$ \\
\hline
ER2                         & $[1,10]$                       & $[1,3]$                          & $^{8}$B                         & 1.11                       & $81.6 \pm 0.4$ \\
$E \in [5,16]$\,MeV         &                                        &                                          & $^{8}$Li                        & 1.21                       & $67.5 \pm 0.4$ \\
\hline
\end{tabular}}
\end{center}
\caption{Selection cuts of the candidate events in energy ($E$) and
  time ($t_g$, $t_E$) for the two regimes ER1 and ER2 for isotopes
  {$^{8}$B}, {$^{6}$He} and {$^{8}$Li}. In addition,
  the expected cosmogenic isotopes are given with their lifetimes and
  energy cut efficiencies.
For each energy
regime, the time profile is constructed from events within the time interval
$t_g$ relative to preceding muons. 
To enhance the 
signal-to-background ratio, the energy distribution is based on events
in the time interval $t_E$. 
The distributions and the result of a simultaneous fit in time and energy
to both regimes is shown in figure~\ref{fig:8B_6He_and_8Li}.} 
\label{tab:8B_6He_and_8Li}
\end{table}

\begin{figure}
\centering
\includegraphics[width=0.7\textwidth]{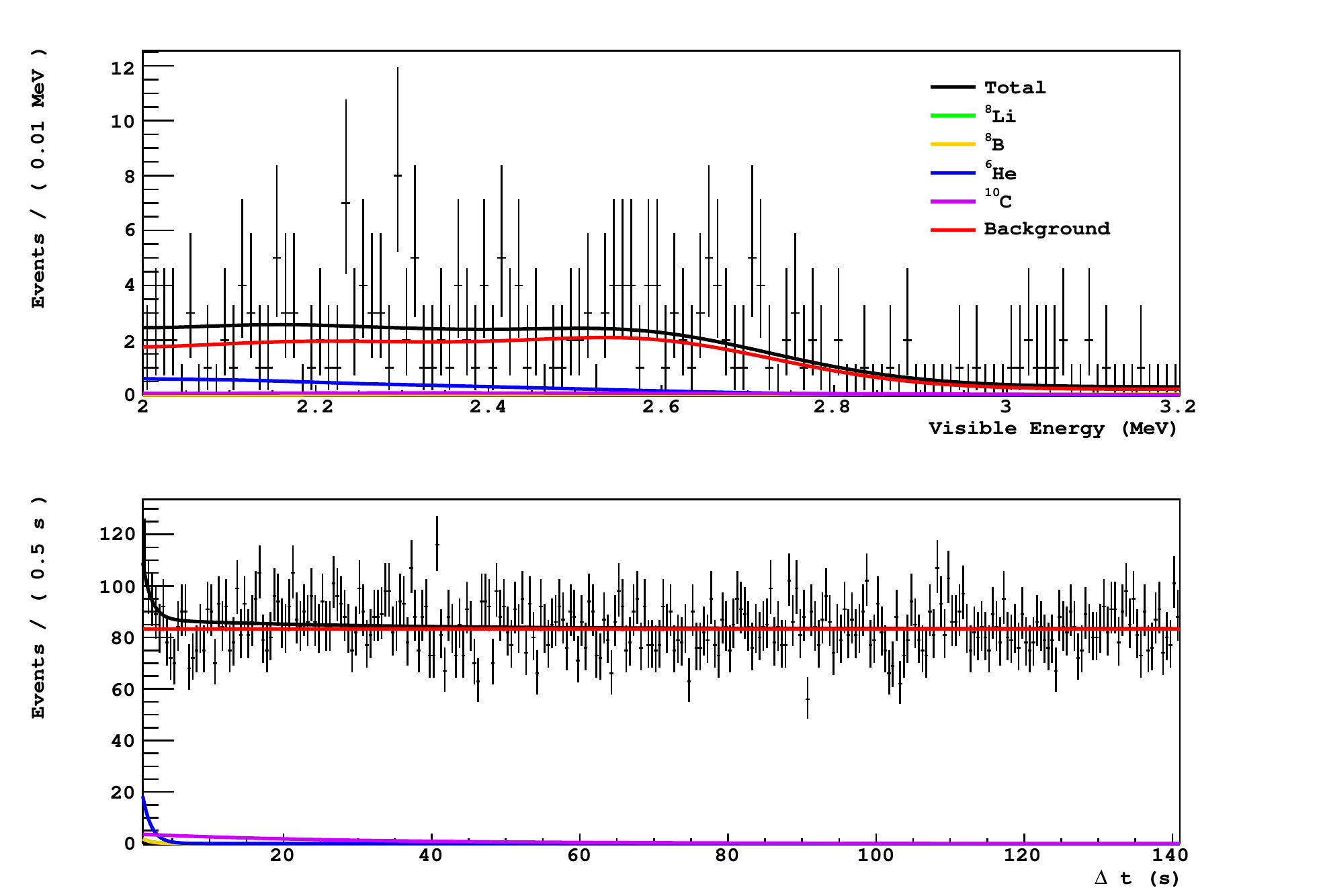}
\includegraphics[width=0.7\textwidth]{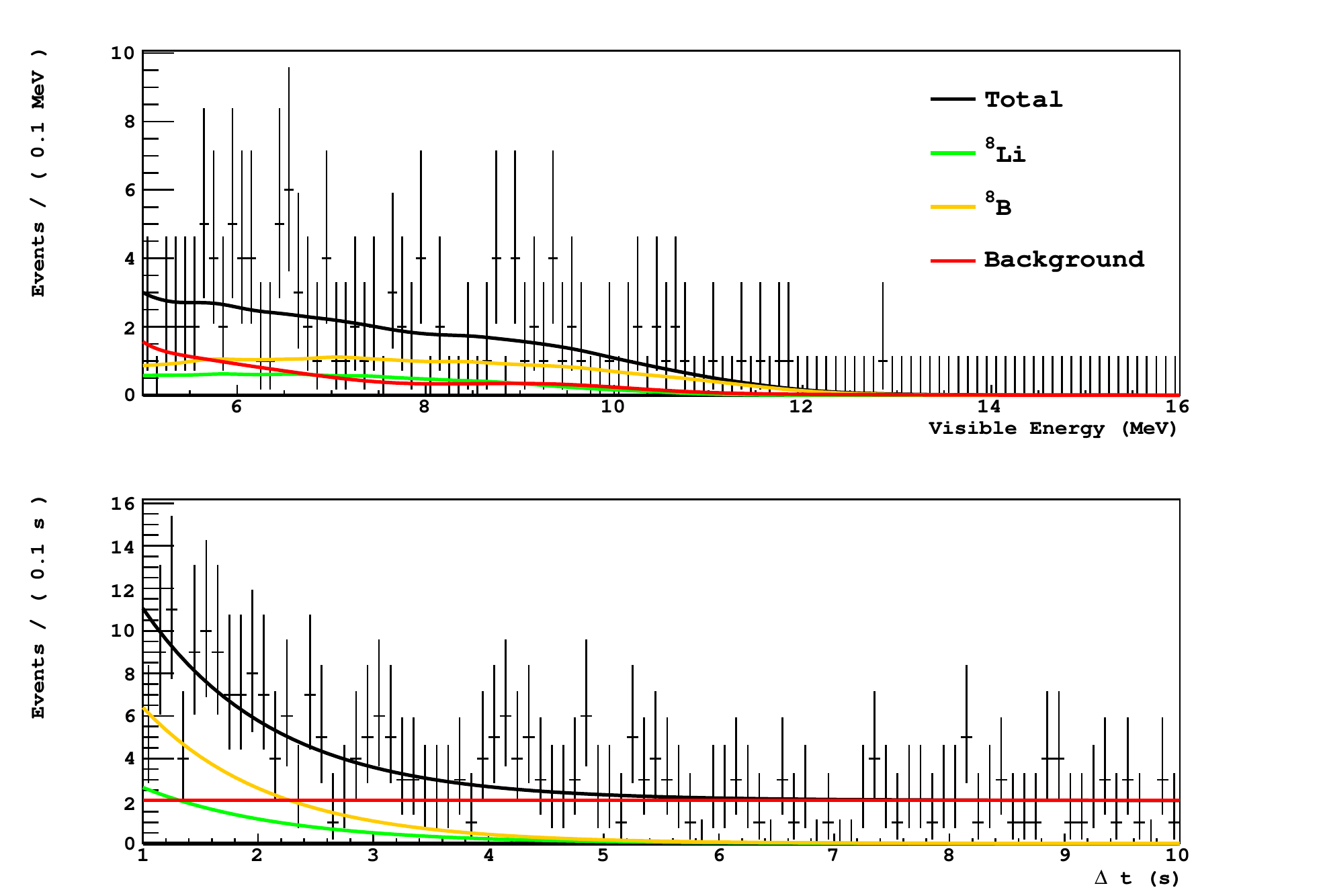}
\caption{Simultaneous fit of the cosmogenic isotopes $^{8}$B,
  $^{6}$He and $^{8}$Li in visible energy deposition
  (first and third panels) and decay time relative to preceding muons (second
  and fourth panels) 
for the energy regimes ER1 ($E\in [2,3.2]$\,MeV) and ER2 ($E\in
  [5,16]$\,MeV). The isotope  
$^{10}$C is included as contaminant. 
Some of the isotopes cannot be seen in the graph because the fit
  returns a very low value for their rates. 
The goodness of the simultaneous fit is $\chi^2/\textrm{ndf} = 457/499$.} 
\label{fig:8B_6He_and_8Li}
\end{figure}

The cosmogenic isotopes $^{8}$B ($\beta ^+$-emitter, $\tau = 1.11$\,s,
$Q = 18.0$\,MeV), $^{6}$He ($\beta ^-$-emitter, $\tau = 1.16$\,s, $Q =
3.51$\,MeV) and $^{8}$Li ($\beta ^-$-emitter, $\tau = 1.21$\,s, $Q =
16.0$\,MeV) feature similar lifetimes.  However, the significantly lower
$Q$-value of $^{6}$He enables a partial disentanglement of these
radionuclides via cuts in visible energy and time. We separate the
energy range in two regimes, denoted as ER1 ($E\in [2,3.2]$\,MeV) 
and ER2 ($E\in [5,16]$ \,MeV), respectively. Regime ER1 comprises
decays of all three isotopes, whereas ER2 includes only {$^{8}$B} and
{$^{8}$Li}. The two energy intervals are fit
simultaneously with their respective time profiles in a single,
un-binned maximum likelihood fit. The spectral shape of uncorrelated
background is derived from events with $\Delta t > 140$\,s. Table~\ref{tab:8B_6He_and_8Li}
summarizes the energy selection efficiencies
and the chosen time gates for the time and energy distributions in these
two energy regimes. Due to the lower energy threshold of ER1, an
additional contribution of the cosmogenic isotope {$^{10}$C} ($\beta
^+$-emitter, $\tau = 27.8$\;s, $Q = 3.65$\,MeV) is included as a free
parameter in the fit. To avoid contaminations of short-living
cosmogenic isotopes, a 1\,s veto after each muon is applied for both
regimes, inducing a dead time of 3.6\,\%. 
The result of the simultaneous fit is shown in
figure~\ref{fig:8B_6He_and_8Li} for the energy regimes ER1 and ER2,
respectively. The isotope production rates are found to be $R(^{8}$B$)
= (0.41\pm 0.16_\textrm{stat} \pm 
0.03_\textrm{syst})$\,(d\,100t)$^{-1}$, $R(^{6}$He$) = (1.11 \pm
0.45_\textrm{stat} \pm 0.04_\textrm{syst})$\,(d\,100t)$^{-1}$ and
$R(^{8}$Li$) = (0.21\pm 0.19_\textrm{stat} \pm
0.02_\textrm{syst})$\,(d\,100t)$^{-1}$. The corresponding yields are
$Y(^{8}$B$) = (1.4 \pm 0.6_\textrm{stat} \pm 0.1_\textrm{syst})\cdot
10^{-6}\,/(\mu \cdot (\rm{g/cm}^{2}))$, $Y(^{6}$He$) = (3.80 \pm
1.53_\textrm{stat} \pm 0.14_\textrm{syst})\cdot 10^{-6}\,/(\mu \cdot
(\rm{g/cm}^{2}))$ and $Y(^{8}$Li$) = (7.1 \pm 6.6_\textrm{stat} \pm
0.7_\textrm{syst})\cdot 10^{-7}\,/(\mu \cdot (\rm{g/cm}^{2}))$.

\subsection{{$^{9}$C}}
\label{subsec::9C}

\begin{figure}
\begin{center}
\includegraphics[width=0.7\textwidth]{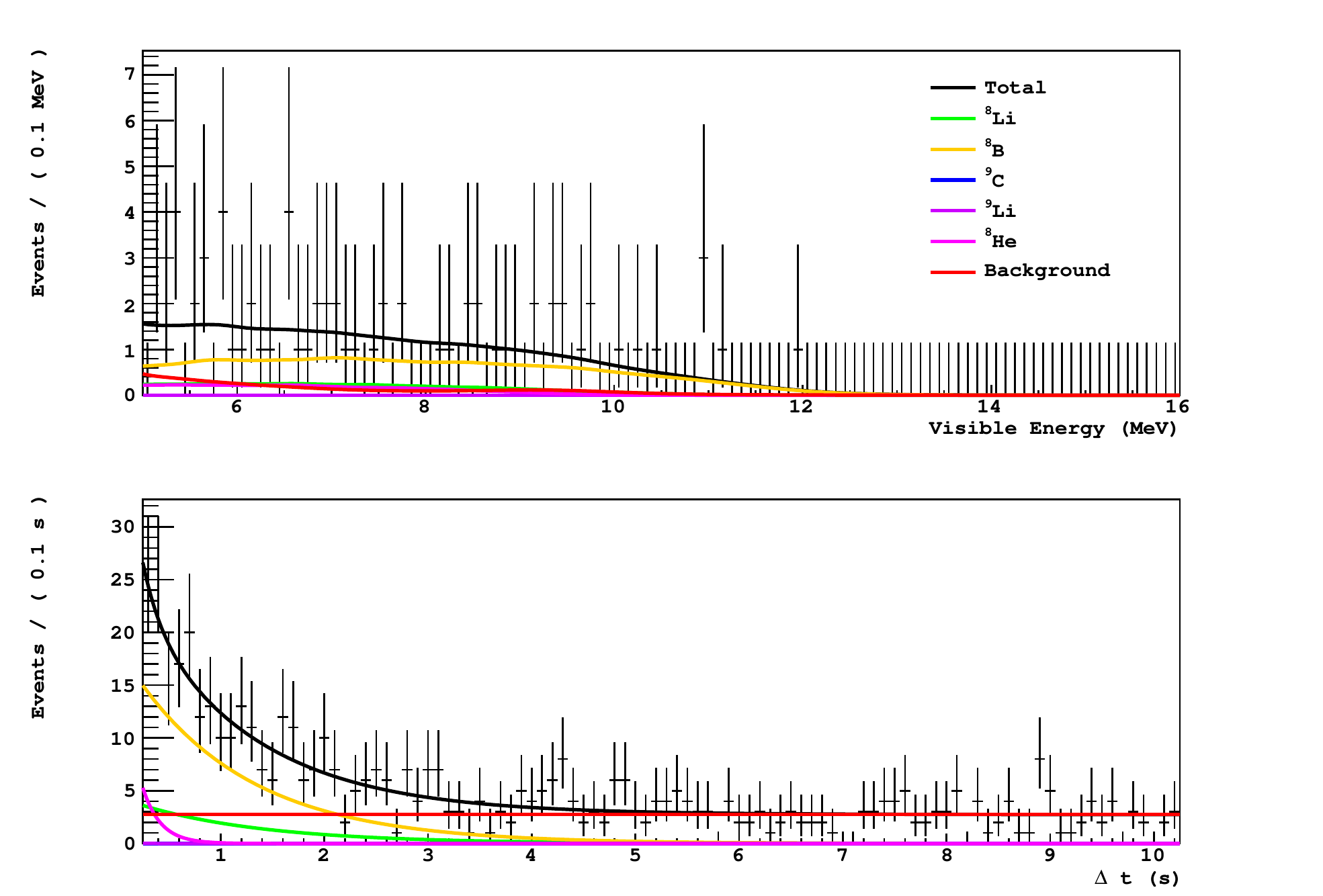}
\caption{Simultaneous fit of the cosmogenic isotope $^{9}$C in
  visible energy deposition (top panel) and decay time relative to preceding
  muons (bottom panel). The fit returns only an upper limit for the isotope
  $^{9}$C and some isotopes which cannot be seen in the graph.
The goodness of the simultaneous fit is $\chi^2/\textrm{ndf} = 218/268$.} 
\label{fig:9C}
\end{center}
\end{figure}

Candidate events for the decay of $^{9}$C ($\beta ^+$-emitter, $\tau =
182.5$\,ms, $Q = 16.5$\,MeV) are selected by $E\in[5,18]$\,MeV and
$\Delta t \in[250\,\rm{ms},10.25\,\rm s]$. The lower energy threshold
avoids decays of $^{6}$He in the 
data set. Events occurring for $\Delta t \in [250,600]$\,ms are
employed to obtain the energy distribution. After each
muon, a $250$\,ms veto is applied, rejecting contributions from
shorter-lived cosmogenic radionuclides and 
reducing the live time of the data set by 0.9\,\%. The isotopes
$^{8}$He, $^{9}$Li, $^{8}$B and $^{8}$Li are treated as
contaminants. Events for $\Delta t > 10.25$\,s are used to build the
spectral shape of the uncorrelated background. 
The energy distribution for the $^{8}$B, $^{6}$He and $^{8}$Li analysis 
(ER2  with $E\in [5,16]$\,MeV, $t_E \in [1,3]$\,s) is used to confine the rates of $^{8}$B
and $^{8}$Li in the simultaneous fit as additional
complementary information (section~\ref{subsec::8B_6He_and_8Li}). 

The result of the best fit for the time profile and both energy distributions
is shown in figure~\ref{fig:9C}. A fraction of 
$\varepsilon(^{9}$C$) = (73.4 \pm 0.4)\,\%$ of all $^{9}$C candidates is 
expected within the energy range of the $^{9}$C
candidates.  The rate and yield of $^{9}$C
are determined to an upper limit of $R(^{9}$C$) <
0.47$\,(d\,100t)$^{-1}$, and $Y(^{9}$C$) < 1.6\cdot 10^{-6}\,/(\mu
\cdot (\rm{g/cm}^{2}))$ at $3\sigma$ confidence level after correcting for 
efficiencies. 

\subsection{{$^{11}$Be}}
\label{subsec::11Be}

\begin{figure}
\begin{center}
\includegraphics[width=0.7\textwidth]{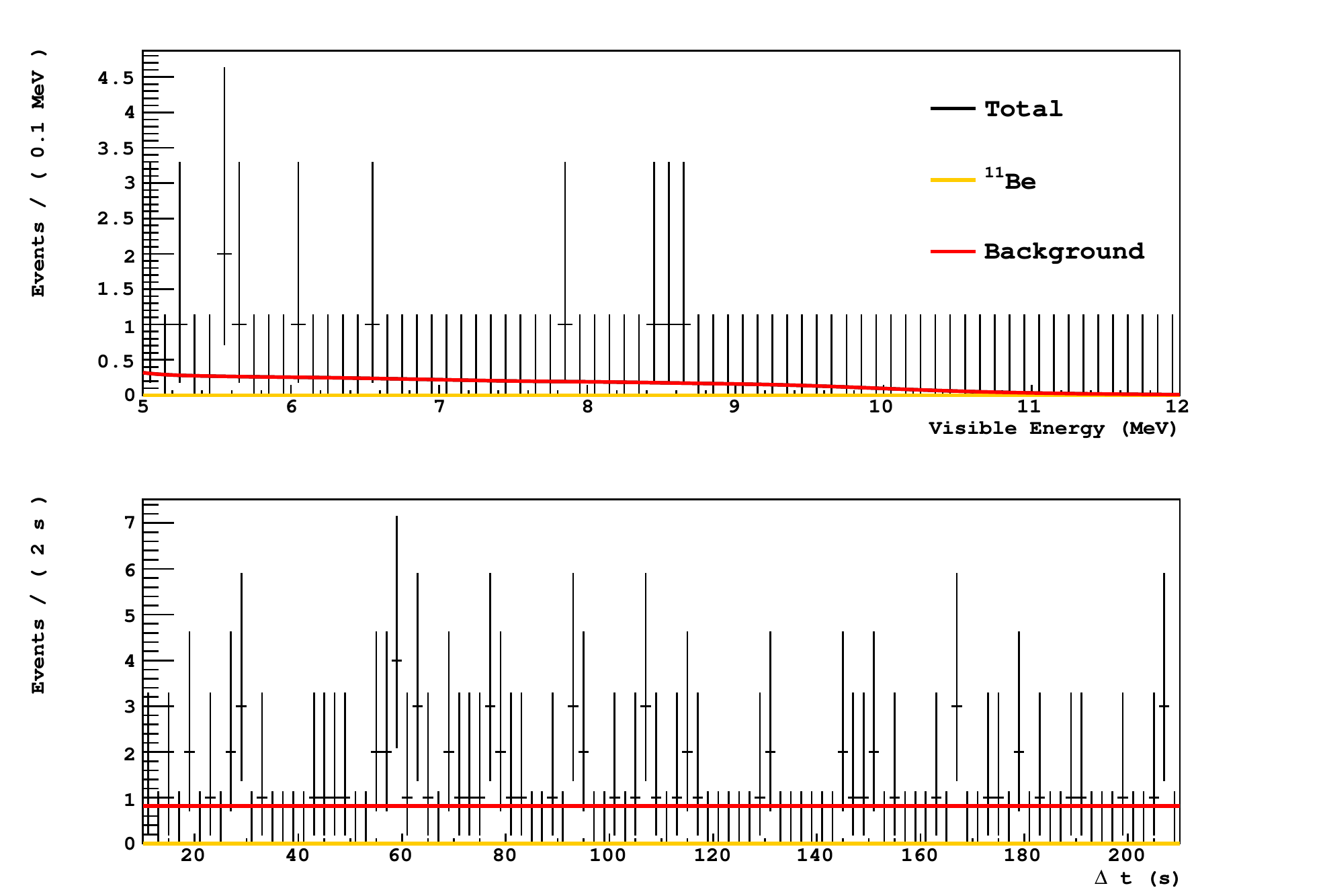}
\caption{Simultaneous fit of the cosmogenic isotope $^{11}$Be
  in visible energy deposition (top panel) and decay time relative to
  preceding muons (bottom panel). The fit returns only an upper limit for
  the isotope $^{11}$Be. The goodness of the simultaneous fit is
  $\chi^2/\textrm{ndf} = 30/61$. We note, that the low value of
  the reduced $\chi^2$ is connected to the low statistics of the data
  set fit in the unbinned maximum likelihood fit.}
\label{fig:11Be}
\end{center}
\end{figure}

Events with $E\in[5,12]$\;MeV are
chosen in order to determine production rate for {$^{11}$Be} 
which is a $\beta ^-$-emitter, with $\tau =
19.9$\,s, $Q = 11.5$\,MeV. The lower energy boundary is needed in order to reject {$^{10}$C} and {$^{11}$C} decays as well as the external (uncorrelated) $\gamma$-background from $^{208}$Tl. The time profile is constructed for
$\Delta t \in [10,210]\;$s with respect to a preceding muon whose track 
is reconstructed
within $1.5$\,m from the {$^{11}$Be} candidate decay.
The energy distribution is composed of events with $\Delta
t \in [10,40]$\;s in order to
increase the signal-to-background ratio after applying the same muon track 
cut. A 10\,s veto
after each muon rejects the shorter-lived cosmogenic radionuclides, which
decreases the live time of the data set by $28.4$\,\% and leaves
{$^{11}$Be} as the only cosmogenic isotope. The background spectral
shape is derived from events which satisfy the muon track cut and
occur later than 210\,s after the muon. 
The best fit results are shown in figure~\ref{fig:11Be}. The selection
efficiency of the muon track cut is estimated to $(63.3 \pm 2.5)$\,\%
using the time and lateral distribution of cosmogenic {$^{12}$B} to
preceding muon tracks. The fraction 
of decays in the selected visible energy range is calculated to
to be $\varepsilon(^{11}\textrm{Be}) = (69.3 \pm 0.6)\,\%$. The isotopic production
rate and yield of {$^{11}$Be} are measured to be $R(^{11}\textrm{Be}) <
0.20$\,(d\,100t)$^{-1}$ and $Y(^{11}$Be$) < 
7.0\cdot 10^{-7}\,/(\mu \cdot (\rm{g/cm}^{2}))$ at the $3\sigma$ confidence 
level.

\subsection{{$^{10}$C}}
\label{subsec::10C}

\begin{figure}
\begin{center}
\includegraphics[width=0.7\textwidth]{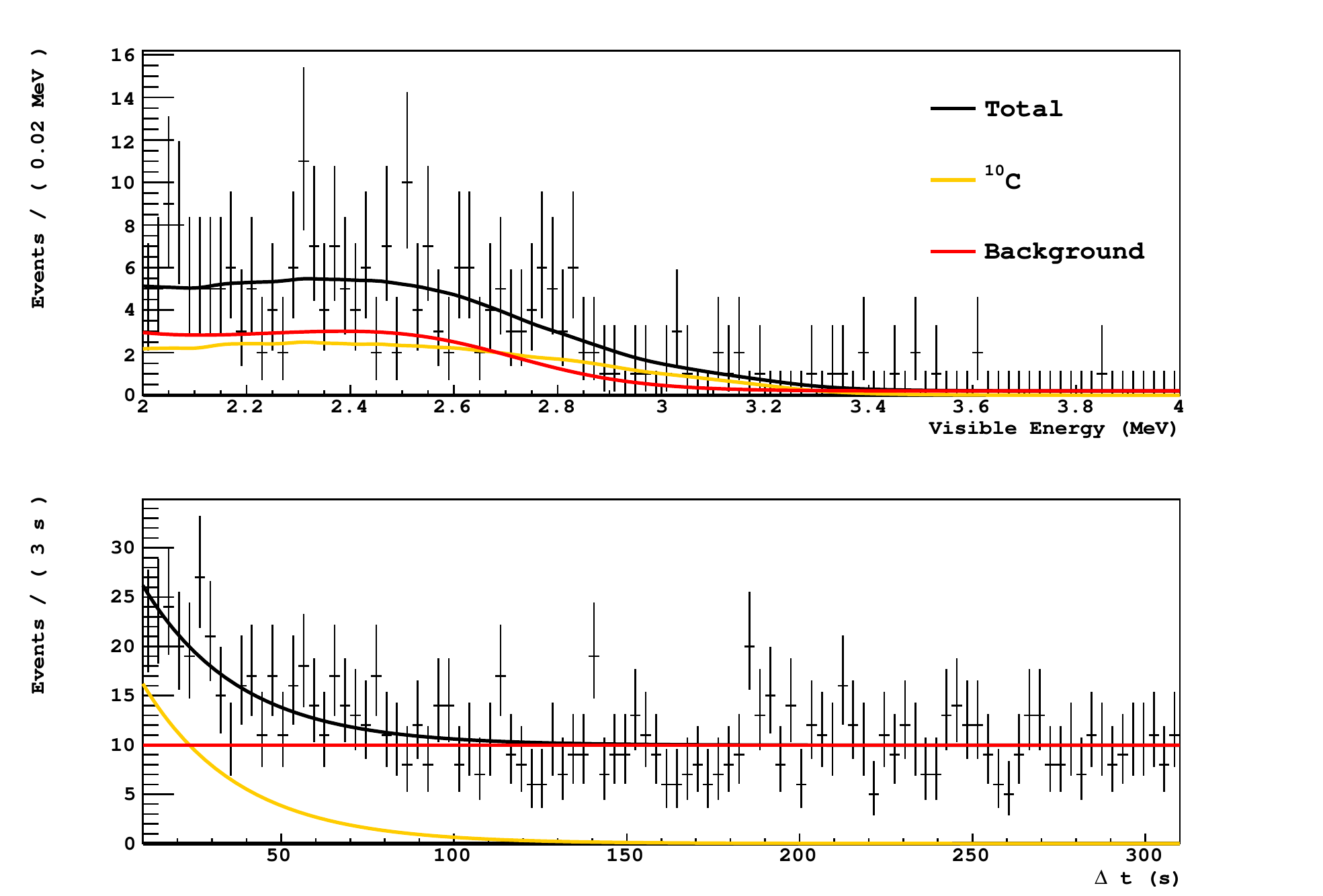}
\caption{The simultaneous fit of the cosmogenic isotope $^{10}$C
  in visible energy deposition (top panel) and decay time relative to
  preceding muons (bottom panel). The goodness of the simultaneous fit is
  $\chi^2/\textrm{ndf} = 153/162$.}  
\label{fig:10C}
\end{center}
\end{figure}

The production of {$^{10}$C} ($\beta ^+$-emitter, $\tau = 27.8$\,s, $Q
= 3.65$\,MeV) in muon-induced spallation processes on $^{12}$C is
usually accompanied by the emission of at least one free neutron. These
neutrons are eventually captured 
on hydrogen or carbon
(section~\ref{subsec::8He_and_9Li}). The accidental background is 
significantly reduced after requiring a three-fold
coincidence between a muon, at least one subsequent neutron capture, and a
{$^{10}$C} decay candidate.
Neutron captures with a minimum energy of $1.3$\;MeV are selected inside
the full neutron trigger gate of $[16,1600]$\;\textmu s after the muon
event. Also {$^{10}$C} candidates must satisfy a cut in visible
energy of $[2,4]$\;MeV and 
occur within $[10,310]$\;s to a preceding $\mu n$-coincidence. A
lower energy threshold of 2\,MeV avoids a contribution of {$^{11}$C}
decays in the data set. The energy distribution of the {$^{10}$C}
candidates is constructed from events with 
$\Delta t \in [10,50]$\;s. Only {$^{11}$Be} contributes as a cosmogenic
contaminant in this parameter selection. Based on the selection cuts
and the additional requirement of a $\mu n$-coincidence, the
contribution of {$^{11}$Be} is estimated 
to be less than $6 \cdot 10^{-3}$\,(d\,100t)$^{-1}$ and this is taken into
account as a systematic uncertainty. The spectral shape of uncorrelated
background is derived from events at $\Delta t >310$\,s. The
simultaneous fit is depicted in 
figure~\ref{fig:10C}. The fraction of {$^{10}$C} decays accompanied by a
muon in coincidence with at least one detected neutron capture is
estimated via a test sample of {$^{10}$C} candidates. {$^{10}$C} 
candidates are selected by $\Delta t \in [10,310]$\;s and a lateral
distance of 1\,m to a parent muon after removal of the
neutron requirement. The numbers of {$^{10}$C} decays in
the subset which satisfy the neutron requirements (subset A), as well as
in the complementary subset 
(subset B), are derived by time profile fits and then compared. The
{$^{10}$C} tagging efficiency due to the neutron requirement is
evaluated to $\varepsilon_{n}(^{10}\rm C) =
(92.5^{+7.5}_{-20.0})\,\%$. The broad uncertainty range is associated 
with the determination of {$^{10}$C} decays in subset B, which is
dominated by physically uncorrelated matches. Based on the Monte Carlo
simulation, we expect $\varepsilon(^{10}\rm C) = (79.0 \pm 0.5)\,\%$ of
all {$^{10}$C} decays within the 
selected energy range. By considering these corrections, the
{$^{10}$C} rate and yield are determined by the simultaneous fit to
$R(^{10}\rm C) = (0.52 \pm 0.07_\textrm{stat} {}^{+0.11}_{-0.06}
{}_\textrm{syst})$\,(d\,100t)$^{-1}$ and $Y(^{10}$C$) = (1.79 \pm
0.25_\textrm{stat} {}^{+0.38}_{-0.20} {}_\textrm{syst})\cdot
10^{-6}\,/(\mu \cdot (\rm{g/cm}^{2}))$. 

\subsection{{$^{11}$C}}
\label{subsec::11C}

Like {$^{10}$C}, neutron emission is expected in the muon-induced
production of cosmogenic {$^{11}$C} which is a $\beta ^+$-emitter, with $\tau =
29.4$\,min , $Q = 1.98$\,MeV. Therefore, an analogous three-fold
coincidence    between a muon, subsequent neutron capture(s), and {$^{11}$C}
candidates is applied in the rate determination. Candidates of
{$^{11}$C} decays are selected within the energy range of $[1,2]$\;MeV
and the time gate of  $[0.1,3.6]$\,h with respect to a preceding muon-neutron
coincidence. The events for the energy distribution are selected in
the same time interval. 
As a result of the long lifetime of {$^{11}$C} and an average run
duration of $\sim$6\,h in Borexino, effects of run boundaries on the
time profile are not negligible. To avoid a distortion of the time
profile, {$^{11}$C} decays within the 
first $3.6$\,h after run start are not considered in the analysis. This
restriction reduces the data set to a live time of 188\,d. Events within 
2\,h and 4\,m with respect to any neutron capture vertex of a 
$\mu n$-coincidence are vetoed to obtain the
formation of the background spectral shape.  The remaining events are used 
to derive a spectrum containing
only $\mu n$-uncorrelated background sources, i.e. non-cosmogenic
background, and {$^{11}$C} production without the detection of a
subsequent neutron capture. The latter contribution is the result of
the limited neutron detection efficiency, in case of saturated detector
electronics and
so-called \textit{invisible} channels. Invisible channels denote all
muon-induced production processes, yielding {$^{11}$C} with no free
neutron emission in the final state \cite{gal04,Abe:2009aa}.

\begin{figure}
\begin{center}
\includegraphics[width=0.7\textwidth]{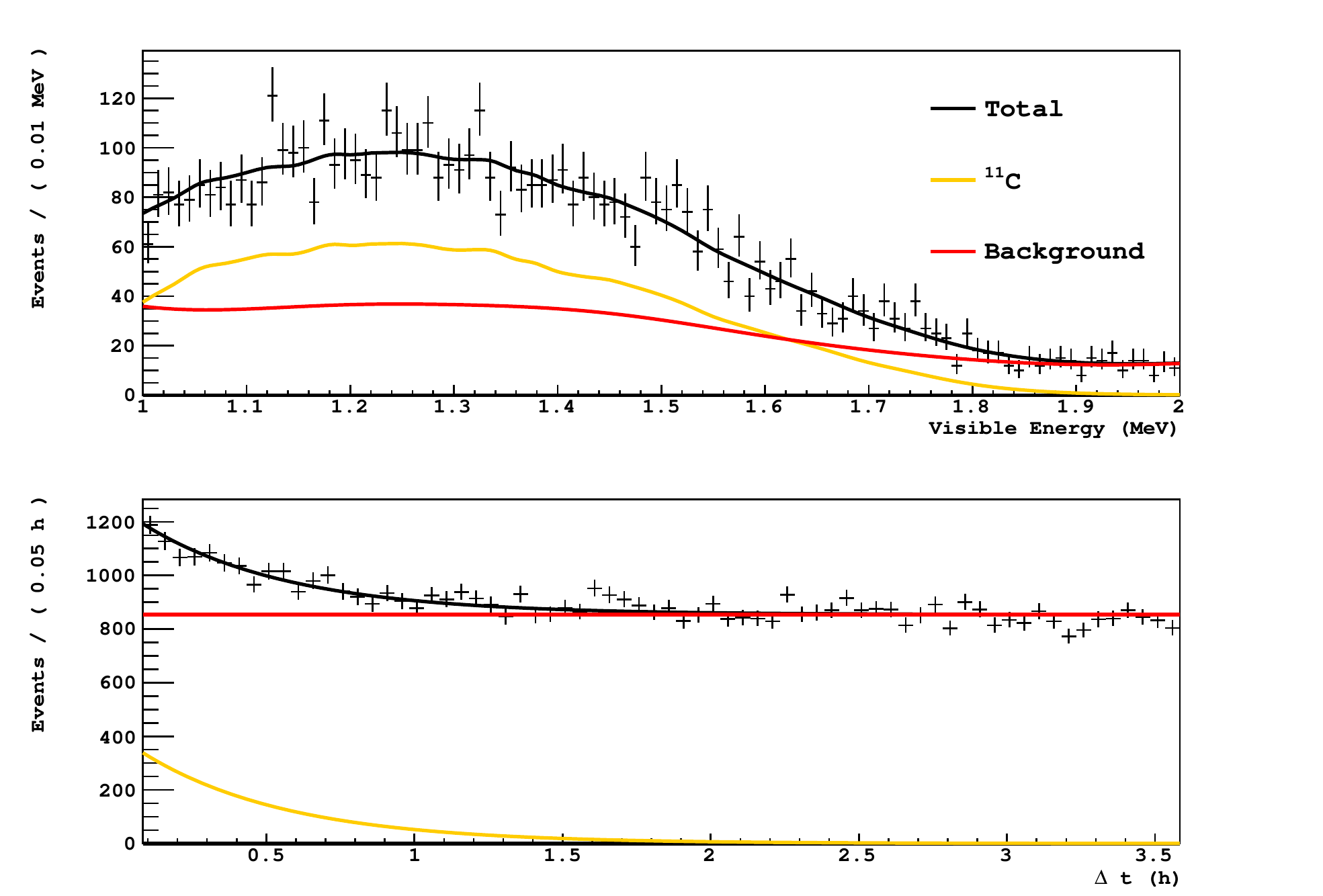}
\caption{Simultaneous fit of the cosmogenic isotope $^{11}$C
  in visible energy deposition (top panel) and decay time relative to
  preceding muons (bottom panel). The goodness of the simultaneous fit is
  $\chi^2/\textrm{ndf} = 189/168$.}  
\label{fig:11C}
\end{center}
\end{figure}

The result of a simultaneous fit in energy and time is shown in
figure~\ref{fig:11C}. The fraction of {$^{11}$C} decays correlated with
a muon and at least one 
detected neutron is estimated in the same manner as in the  {$^{10}$C}
analysis described in section~\ref{subsec::10C}. The selection of the
subset of {$^{11}$C} 
candidates is chosen within $\Delta t \in[0.1,3.6]$\;h and a lateral
distance of 1\,m to a preceding $\mu n$-coincidence. The 
obtained efficiency for  the neutron requirement
is $\varepsilon_{n}(^{11}\rm C) = (86.8 \pm
6.9)\,\%$.  The
cosmogenic production rate is found to be $R(^{11}\rm C) = (25.8 \pm
1.3_\textrm{stat} \pm 
3.2_\textrm{syst})$\,(d\,100t)$^{-1}$ with a corresponding yield of
$Y(^{11}$C$) = (8.86 \pm 0.45_\textrm{stat} \pm
1.10_\textrm{syst})\cdot 10^{-5}\,/(\mu \cdot 
(\rm{g/cm}^{2}))$ after correcting for the fraction of $\varepsilon(^{11}\rm C) =
(92.2 \pm 0.4)\,\%$ of all decays which
deposit a visible energy in the selected parameter space. The rate of 
{$^{11}$C} decays detected in
coincidence with cosmic muons and associated neutrons is in good
agreement with results obtained 
from spectral fits without the coincidence requirement. Modeling all
signal components in the energy range of $[270,1600]$\,keV, the
{$^{11}$C} rate was found to 
be $R(^{11}\rm C) = (28.5 \pm 0.2_\textrm{stat} \pm
0.7_\textrm{syst})$\,(d\,100t)$^{-1}$ in the precision measurement of
the solar {$^{7}$Be} neutrino interaction rate in Borexino
\cite{bx11be7}. 

 \section{Monte Carlo simulations of Cosmogenic Neutrons and Radioisotopes}
 \label{sec::mc}

 The production of cosmogenic neutrons and radioisotopes in Borexino was studied
 by using the \Geant4 \cite{bib:geant4a, bib:geant4b} and
 F{\sc{luka}}~\cite{bib:fluka1,bib:fluka2} simulation packages which are both
 commonly used to simulate deep underground, low background experiments.  Their
 predictions are compared with our experimental data.

 \subsection{Simulation procedure}

 Initially, a careful description of the muon-induced radiation field
 at LNGS was prepared using \Flukadot.  For this simulation a
 model of Hall C is surrounded 
 by a 700\,cm thick shell of Gran Sasso rock \cite{bib:rock} which is found
 sufficient to allow a full shower development.    The setup is subjected to
 residual cosmogenic muons taking into account the muon angular distribution and
 the muon differential energy spectrum as a function of the slant depth and muon
 event multiplicity as measured by the MACRO experiment
 \cite{bib:macro03, bib:macro2}. 
 The adopted muon charge ratio is  $R_\mu=N_{\mu^+}/N_{\mu^-}\simeq 1.38$ as
 measured by the OPERA experiment \cite{bib:opera}.  Cosmogenic muons and
 muon-induced secondaries emerging into Hall C are followed, and all particles
 reaching the Borexino water tank are recorded.   Details of this simulation
 are given in \cite{bib:fluka3}.

 A fraction of 1.5\,\% of the cosmogenic events with multiple muons
 crossing the Borexino detector 
 simultaneously is found from simulation.   Further, 12\,\% of single
 muon events are actually caused 
 by single muons which belong to muon bundles.

 The cosmogenic events which were recorded at the outside of the water
 tank are then used as 
 a source for both
 \Geant4 and \Fluka to simulate the production of cosmogenic neutrons and
 radioisotopes inside the Borexino detector setup. The yields are
 extracted directly by recording 
 neutron captures and residual isotope production, rather than
 simulating the full detector response. 
 The Monte Carlo simulation predictions are compared to the data
 analysis results of the previous sections 
 that include corrections for detection efficiency.

 For the comparison between predictions   
 and experimental data, the following observables are considered: 1) the
 rates for neutron captures 
 and  cosmogenic isotopes production, 2) the neutron capture time,              
 3) the neutron
 capture multiplicity for individual muon events, and 4) 
the lateral distance between
 the neutron capture and the parent muon track.  We also compare the rate
 of muon events for which one or more neutron captures are recorded.

 \subsubsection{\Geant4}
   \label{sec:g4PL}
 
   The version of \Geant4 which is used for this study is
   Geant4-09-06-patch-01  from February 2013. 
   Detailed description of the \Geant4 toolkit is 
available in \cite{bib:geant4}.
 
   Many hadronic models are available in \Geant4 \cite{bib:g4physman},
   and can be used as functions of 
   the particle energy.
   We cover the entire muon energy spectrum at LNGS through a
   combination of different physics inputs. 
   For our study, we use: 
  a) the Quark-Gluon String (QGS) model for proton, neutron, pion and
   kaon interactions with nuclei at 
     kinetic energies above 12\,GeV, completed with the {\it
   Precompound} model for the evaporation 
     phase of the interaction\,; 
  b) the Fritiof model (FTF) for the interaction of highly energetic
   protons, neutrons, pions and 
     kaons starting from 4-5\,GeV, also completed with the {\it Precompound} 
model\,; 
  c) the Bertini cascade (BERT) model, which includes intra-nuclear
   cascade, followed by precompound 
     and evaporation phases of the residual nucleus, for proton,
   neutron, pion and kaon interactions with 
     nuclei at kinetic energies below 9.9\,GeV\,;
  d) the Binary cascade (BIC) model, a data driven intra-nuclear
   cascade model intended for energies 
     below 5\,GeV\,;
  e) the High Precision Neutron (HP) model, describing parameterized
   capture and fission for low-energy 
     neutrons (below 20\,MeV)\,;
  f) the Low Energy Parameterized (LEP) model for proton, neutron,
   pion and kaon interactions with 
     nuclei at kinetic energies between 9.5\,GeV and 25\,GeV.
 
   Some ``ready-made" Physics Lists merging different models are
   available, and we have defined four models: Model I 
   (merge of FTF and BIC with HP manually added), Model II (merge of
   FTF, BERT and HP), 
   Model III (merge of QGS, BIC and HP), Model IV (merge of QGS, BERT and HP). 
   A summary of the introduced hadronic models is shown in
   table~\ref{tab:g4physlist}.  

   For each model, G4MuonNuclearProcess was used to simulate the muon-nuclear interactions.
   Below 10\,GeV, the virtual photon is converted into a real photon and then interacts with the nucleus using the BERT model. Above 10\,GeV,
   the virtual photon is converted into a $\pi^0$ and the interaction with the nucleus is described by the FTF model \cite{bib:g4physman}.

  \begin{table}
  \begin{center}
  {\footnotesize
  \begin{tabular}{|l|cccc|}
  \hline
  \textbf{Model I}    &  \textbf{HP}   & \textbf{Binary}  &
  \textbf{Bertini} &  \textbf{FTF} \\ 
  \hline
  Protons & & 0 $\rightarrow$ 5\,GeV & &  4\,GeV $\rightarrow$ 100\,TeV  \\
  Neutrons &0 $\rightarrow$ 20\,MeV & 19.9\,MeV $\rightarrow$ 5\,GeV &
  &  4\,GeV $\rightarrow$ 100\,TeV  \\ 
  $\pi$ & & 0 $\rightarrow$ 5\,GeV &  &  4\,GeV $\rightarrow$ 100\,TeV \\
  K & &  &0 $\rightarrow$ 5\,GeV  &  4\,GeV $\rightarrow$ 100\,TeV \\
  \hline
  \textbf{Model II}            &        \textbf{HP}            &
  \textbf{Bertini}        &       & \textbf{FTF}  \\ 
  \hline
  Protons &  &  0 $\rightarrow$ 5\,GeV &  &   4\,GeV $\rightarrow$ 100\,TeV  \\
  Neutrons &  0 $\rightarrow$ 20\,MeV&  19.9\,MeV $\rightarrow$ 5\,GeV
  &  &  4\,GeV $\rightarrow$ 100\,TeV  \\ 
  $\pi$, K &  &  0 $\rightarrow$ 5\,GeV &  &  4\,GeV $\rightarrow$ 100\,TeV  \\
  \hline
  \textbf{Model III}     & \textbf{HP}      & \textbf{Binary}      &
  \textbf{LEP} &  \textbf{QGS} \\ 
  \hline
  Protons & &  0 $\rightarrow$ 9.9\,GeV & 9.5 $\rightarrow$ 25\,GeV &
  12\,GeV $\rightarrow$ 100\,TeV \\ 
  Neutrons & 0 $\rightarrow$ 20\,MeV &  19.9\,MeV $\rightarrow$
  9.9\,GeV & 9.5 $\rightarrow$ 25\,GeV & 12\,GeV $\rightarrow$
  100\,TeV\\ 
  $\pi$, K & &  0 $\rightarrow$ 9.9\,GeV & 9.5 $\rightarrow$ 25\,GeV &
  12\,GeV $\rightarrow$ 100\,TeV \\ 
  \hline
  \textbf{Model IV}     & \textbf{HP }     & \textbf{Bertini}      &
  \textbf{LEP} &  \textbf{QGS} \\ 
  \hline
  Protons & &  0 $\rightarrow$ 9.9\,GeV & 9.5 $\rightarrow$ 25\,GeV &
  12\,GeV $\rightarrow$ 100\,TeV \\ 
  Neutrons & 0 $\rightarrow$ 20\,MeV &  19.9\,MeV $\rightarrow$
  9.9\,GeV & 9.5 $\rightarrow$ 25\,GeV & 12\,GeV $\rightarrow$
  100\,TeV\\ 
  $\pi$, K & &  0 $\rightarrow$ 9.9\,GeV & 9.5 $\rightarrow$ 25\,GeV &
  12\,GeV $\rightarrow$ 100\,TeV \\ 
  \hline
  \end{tabular}}
  \end{center}
  \caption{Summary of the Hadronic Models used in \Geant4.}
  \label{tab:g4physlist}
  \end{table}
 
   Each model has been tested with and without the activation of the
   Light Ion (LI)  Physics List (which defines the light ions likely 
   to be produced by the hadronic interactions, such as deuterons,
   tritons, $^3$He, $\alpha$-particles and generic ions). 
   In addition, the following Physics Lists are included for each
   model: the Electromagnetic Processes for Leptons 
   (G4eMultipleScattering, G4eIonisation, G4eBremsstrahlung, ...), the
   Nuclear Decay Processes (G4Decay, G4Radioac\-tiveDecay) and the 
   standard Elastic Scattering for hadrons (G4HadronElasticPhysics).

  \subsubsection{\Fluka}

 \Fluka is a fully integrated particle physics Monte Carlo simulation package based 
 predominantly on original and well-tested microscopic models. The models are benchmarked 
 and optimized by comparing to experimental data at the single interaction level.
 The physics models in \Fluka are fully integrated into the code and no modifications
 or adjustments are available at the user level.
 A list of benchmark results relevant to the simulation of deep
 underground cosmogenic backgrounds is described in \cite{bib:fluka3}.

 Details of the physics models implemented in \Fluka with focus on hadronic interactions 
 and the \Fluka specific nuclear interaction model PEANUT can be found in
 \cite{bib:fluka4,bib:fluka5,bib:fluka6,bib:fluka7}, while a description of the
 approach for muon interactions in \Fluka is given in \cite{bib:NY_fluka}.
 A validation of the \Fluka Monte Carlo code for predicting induced radioactivity is
 given, for instance, in \cite{bib:fluka_activation}.

 In general, the simulation was performed using the \Fluka default setting PRECISIO(n).
 In addition, photonuclear interactions were
 enabled through the \Fluka option PHOTONUC and a more detailed treatment of nuclear
 de-excitation was requested with the EVAPORAT(ion) and COALESCE(nce) options.
 These enable the evaporation of heavy fragments ($A>1$) and the emission of
 energetic light fragments, respectively. The treatment
 of nucleus-nucleus interaction was turned on for all energies via the option
 IONTRANS and radioactive decays were activated through the option RADDECAY.

 The version of \Fluka used for the present study is FLUKA2011.2, released in November 2011. 
 Further information about the implemented physics models is available through the \Fluka 
 manual and additional documentation and lecture notes located at the
 official \Fluka  website~\cite{bib:fluka0}.

 \subsection{Simulation results}

We present the predictions obtained with \Geant4 and \Fluka regarding
different physics observables.  
For \Geant4, we concluded that the best match to data is given
by Model III or IV, depending on the observable under study. 
Model I produced 15\,\% less neutrons than the other models and Model
II shows no relevant difference with respect to model IV.  
For readability we choose to present here only Model III and IV,
however complete results for all four models are available as supplementary material of this article.

 \subsubsection{Cosmogenic Radioisotopes}

 Production yields for cosmogenic isotopes and neutrons are summarized
 in table~\ref{tab:g4_li}. 
 The yields measured by Borexino are compared to \Geant4 and \Fluka
 predictions as well as measured yields from the KamLAND
 experiment~\cite{Abe:2009aa} given in the rightmost column.  
We note that KamLAND has a different number of carbon nuclei per ton
 of liquid scintillator 
(4.30$\cdot$10$^{28}$ for KamLAND as opposed to 4.52$\cdot$10$^{28}$
 in case of Borexino) and that the mean residual muon energies
 differ somewhat between the two sites: (283\,$\pm$\,19)\,GeV at
 LNGS\footnote{ The residual mean muon energy is based on the MACRO
 measurement for single and double muon events reported in
 \cite{bib:macro3}.  Details of the procedure are given in
 \cite{bib:fluka3}.} versus (260\,$\pm$\,8)\,GeV at the Kamioka
 mine \cite{Abe:2009aa}, which should lead to a difference in the
 observed production yields.

\begin{table}
\begin{center}
\begin{tabular}{l  R{7mm}@{\hspace{2pt}}R{3mm}@{\hspace{2pt}}L{8mm} R{7mm}@{\hspace{2pt}}C{3mm}@{\hspace{2pt}}L{8mm} R{7mm}@{\hspace{2pt}}C{3mm}@{\hspace{2pt}}L{8mm} R{7mm}@{\hspace{2pt}}C{3mm}@{\hspace{2pt}}L{8mm} R{14mm}@{\hspace{2pt}}C{3mm}@{\hspace{2pt}}l }
\hline
& \multicolumn{3}{c}{\bf{\Geant4}}    & \multicolumn{3}{c}{\bf{\Geant4}}    & \multicolumn{3}{c}{\bf{\Fluka}}  & \multicolumn{3}{c}{\bf{Borexino}}  & \multicolumn{3}{c}{\bf{KamLAND}}  \\
& \multicolumn{3}{c}{Model III} & \multicolumn{3}{c}{Model IV}\\
&\multicolumn{12}{c}{--- \small{$\mathrm{\left< E_\mu \right>=283\,\pm\,19\,GeV }$} ---}&
\multicolumn{3}{c}{\small{$\mathrm{\left< E_\mu \right>=260\,\pm\,8\,GeV }$ }}\\
\hline
\\[-10pt]
\bf{Isotopes}&\multicolumn{15}{c}{Yield\ \, $[10^{-7}\,(\mu\,\textrm{g}/\textrm{cm}^2)^{-1}]$}\\
            {\bf $^{12}$N } & 1.11 &$\pm$& 0.13 &  3.0 &$\pm$& 0.2  &  0.5 &$\pm$& 0.2  &   &$<$& 1.1     & 1.8 &$\pm$& 0.4 \\%d
            {\bf $^{12}$B } & 30.1 &$\pm$& 0.7  & 29.7 &$\pm$& 0.7  & 28.8 &$\pm$& 1.9  &  56 &$\pm$& 3   &  42.9 &$\pm$& 3.3   \\%d
{\bf \hspace{4pt}$^{ 8}$He} & &$<$& 0.04 & 0.18 &$\pm$& 0.05 & 0.30 &$\pm$& 0.15 &   &$<$& 1.5    & 0.7 &$\pm$& 0.4 \\%d
{\bf \hspace{4pt}$^{ 9}$Li} & \hspace{1pt}~0.6  &$\pm$& 0.1  & 1.68 &$\pm$& 0.16 &  3.1 &$\pm$& 0.4  & 2.9 &$\pm$& 0.3 & 2.2 &$\pm$& 0.2 \\%d
{\bf \hspace{4pt}$^{ 8}$B } & 0.52 &$\pm$& 0.09 & 1.44 &$\pm$& 0.15 &  6.6 &$\pm$& 0.6  &  14 &$\pm$& 6   & 8.4 &$\pm$& 2.4 \\
{\bf \hspace{4pt}$^{ 6}$He} & 18.5 &$\pm$& 0.5  &  8.9 &$\pm$& 0.4  & 17.3 &$\pm$& 1.1  &  38 &$\pm$& 15  &\multicolumn{3}{c}{not reported}  \\
{\bf \hspace{4pt}$^{ 8}$Li} & 27.7 &$\pm$& 0.7  &  7.8 &$\pm$& 0.4  & 28.8 &$\pm$& 1.0  &  7  &$\pm$& 7   &  12.2 &$\pm$& 2.6   \\
{\bf \hspace{4pt}$^{ 9}$C } & 0.16 &$\pm$& 0.05 & 0.99 &$\pm$& 0.13 & 0.91 &$\pm$& 0.10 &   &$<$& 16      & 3.0 &$\pm$& 1.2 \\
            {\bf $^{11}$Be} & 0.24 &$\pm$& 0.06 & 0.45 &$\pm$& 0.09 & 0.59 &$\pm$& 0.12 &   &$<$& 7.0     & 1.1 &$\pm$& 0.2 \\%d
            {\bf $^{10}$C } & 15.0 &$\pm$& 0.5  & 41.1 &$\pm$& 0.8  & 14.1 &$\pm$& 0.7  &  18 &$\pm$& 5   &  16.5 &$\pm$& 1.9   \\%d
            {\bf $^{11}$C } & 315  &$\pm$& 2    &  415 &$\pm$& 3    &  467 &$\pm$& 23   & 886 &$\pm$& 115 & 866 &$\pm$& 153 \\%d
\hline
\\[-10pt]
\textbf{Neutrons}& \multicolumn{15}{c}{Yield\ \, $[10^{-4}\,(\mu\,\textrm{g}/\textrm{cm}^2)^{-1}]$}\\
               & 3.01 &$\pm$& 0.05 & 2.99 &$\pm$& 0.03 & 2.46 &$\pm$& 0.12 & 3.10 &$\pm$& 0.11 & 2.79 &$\pm$& 0.31\\
\hline
\end{tabular}
\end{center}
\caption{Predicted yields for cosmogenic products obtained from
         \Geant4 (Model III and IV) and \Fluka are compared to data
         from Borexino . 
         Also shown are results from the KamLAND experiment
         \cite{Abe:2009aa}. Note that the production yields depend on
         the number of carbon 
atoms per weight and the muon energy spectrum. 
         Thus, a 10\,--\,20\,\% difference between KamLAND and Borexino
         results is expected.}
\label{tab:g4_li}
\end{table}

\subsubsection{Cosmogenic neutrons}
\label{sec:sim_cosmo_neut}

\paragraph{Neutron capture time.}
 The simulated neutron capture time of the Borexino scintillator from \Geant4   %254.9 /pm 0.6
 and \Fluka are (275.8\,$\pm$\,0.9)\,\textmu s\footnote{The out-dated
\Geant4 version 4.9.2.p02 returns (254.9\,$\pm$\,0.6)\,\textmu s and is thus in agreement with the measured value.
No explanation has been found for the discrepancy between the different \Geant4 versions.} and
 (253.4\,$\pm$\,0.6)\,\textmu s, respectively.  This is to be compared to the
 measured capture time of 
  (259.7\,$\pm$\,1.3$_\textrm{stat}$\,$\pm$\,2.0$_\textrm{syst}$)\,\textmu s.  
The neutron capture time was also measured in Borexino using an Am-Be
 neutron source \cite{bx11muo} which yields (254.5\,$\pm$\,1.8)\,\textmu s.  
The experimental disagreement with the value measured from cosmogenic neutrons 
could be explained by a fraction of neutrons which are captured on iron in
 the source capsule.  This was also observed by KamLAND  \cite{Abe:2009aa}. 

\paragraph{Neutron production yield.}
In table~\ref{tab:g4_li}, the neutron production yield is reported.
 The observed neutron production deficit of the \Fluka simulation was
 studied in \cite{bib:fluka3}.  The main cause 
 of the deficit was found to be the low cosmogenic production rate
 predicted for $^{11}$C (table~\ref{tab:g4_li}). 
At the LNGS depth, the production of $^{11}$C in liquid scintillator
 is followed by a 
 neutron emission in 95\,\% of all cases as was shown by \cite{gal04}.  
 Since the measured $^{11}$C rate is almost 30\,\% of the neutron
 production rate, and the $^{11}$C rate given by \Fluka is roughly
 50\,\% of the measured value, a reduction of the number of predicted
 cosmogenic neutrons in the order of 15\,\% is expected. 
 The origin of the low $^{11}$C production rate in \Fluka is addressed
 by improvements to the Fermi break-up 
 model~\cite{bib:fluka9, bib:fluka10} which will be available with the
 next \Fluka release.  The impact of 
 the improved model for the $^{11}$C production in liquid scintillator
 at LNGS energies is currently under 
 investigation.
 In addition, \Fluka predicts the production of energetic deuterons
 ($E_{\rm kin}\,>\,50$\,MeV) inside the liquid 
 scintillator with a rate approximately equal to 4\,\% of the cosmogenic
 neutron production rate.  Energetic deuterons can re-interact 
 with the scintillator leading to a break up (and re-capture) of the
 neutrons. However, these processes are not 
 described by \Fluka as no interaction model is yet implemented for these
 deuterons.  This further reduces the predicted 
 number of cosmogenic neutrons.

The neutron yield of the \Geant4 simulation is in good agreement with
the data, but the $^{11}$C rate is also  $\sim$50\,\% of the
measured value. 
As $^{11}$C is usually produced together with a neutron, this
indicates that the yields of other neutron production channels 
are too high in \Geant4.

 The neutron production yield measured by the KamLAND experiment is
 $(2.79\pm0.31) \cdot
 10^{-4}$\,n/$\mu\cdot(\rm{g/cm}^2)$~\cite{Abe:2009aa}.
 The latest results from the LVD experiment, which is also located at LNGS, 
 indicate a neutron yield of $(2.9\pm0.6)\cdot 10^{-4}$\,n/$\mu\cdot(\rm{g/cm}^2)$~\cite{bib:lvd2013}.

\begin{figure}
\centering
\includegraphics[width=0.8\textwidth]{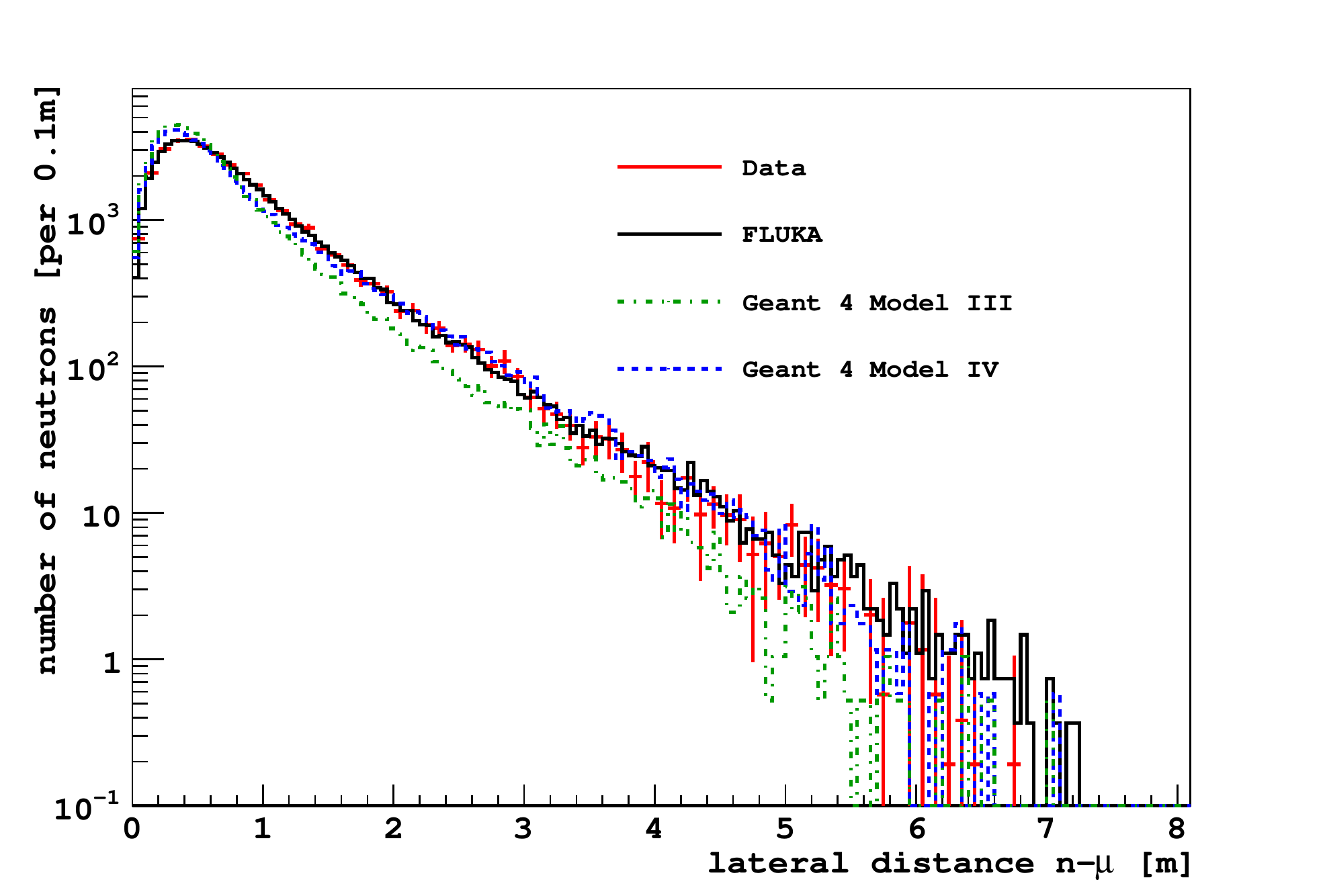}
\caption{Lateral distance between neutron capture points and the
  parent muon track: comparison of Borexino data to
  predictions obtained with \Geant4 - Model III and IV and \Fluka.} 
\label{fig:lateral}
\end{figure}

 \paragraph{Muon-neutron lateral distance distribution.}
 The lateral distance distribution of the neutron capture location
 from the parent muon track is shown in 
 figure~\ref{fig:lateral} and compared to predictions from simulation. As described in
 section~\ref{sec::LDneutrons}, a radial cut of less than 4\,m for neutrons and an
 impact parameter cut less than 4\,m for muons were applied in the
 simulations. 
 Moreover, the muon track reconstruction uncertainties were applied
 \emph{a posteriori} to the simulated distribution. 
 They dominate the shape at small distances. 
 The simulated distributions were scaled to match the number of
 measured neutron captures in order to compare the shape 
 of the histograms.
 The experimental distribution is well reproduced by both
 \Geant4 - Model IV and \Fluka out to large distances. 
\Geant4 - Model III instead reproduces data less accurately.
 The small differences present at the far range can be attributed to
 the additional cuts described in 
 section~\ref{sec::LDneutrons}, which have not been implemented in the
 simulation.  They mostly suppress mis-reconstructed 
 neutrons following showering muons which are expected to dominate the
 distribution at large distances from the track.

\begin{figure}
\centering
\includegraphics[width=0.8\textwidth]{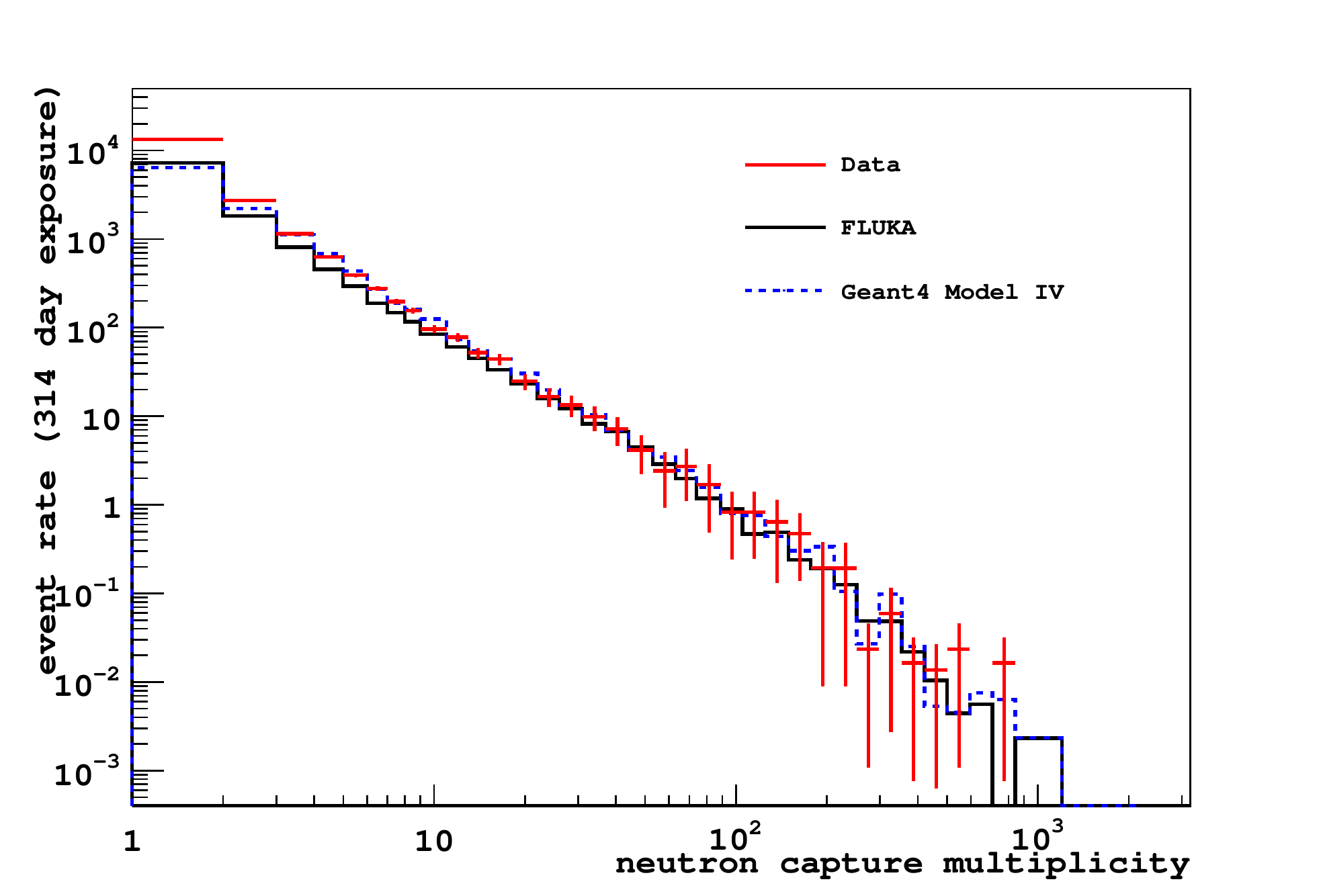}
\caption{Neutron capture multiplicity: comparison of Borexino data
  to predictions obtained with \Geant4 - Model IV and \Fluka.  
Simulated curves were modified to account for the limited neutron
  detection window $[30-1590]$\,\textmu s after the muon.  
\Geant4 - Model III curve does not differ appreciably from Model IV
  curve and is not shown.} 
\label{fig:multiplicity}
\end{figure}

\paragraph{Neutron capture multiplicity.} 
The neutron capture multiplicity distribution is shown in
figure~\ref{fig:multiplicity}, for the the experimental spectrum
as obtained from SYS1 (section~\ref{sec:neutrons_fwfd}). The
experimental distribution is compared 
to predictions by \Geant4 - Model IV and \Fluka 
 which were both scaled to match the live time of the
experimental data set. 
With the exception of events with low multiplicities, good overall
agreement between data and simulation is found for both packages out
to multiplicities of hundreds of neutron captures. 
 The shape of the multiplicity distribution is somewhat distorted towards small
 multiplicities for both Monte Carlo simulations and data with respect to the
 true physical distribution, because neutrons are only detected
 if they are captured between 30 and 1590\,\textmu s after the
 muon trigger.

 \paragraph{Rate of neutron-producing muons.} 
The rate of cosmic muons crossing Borexino which produce at least one
neutron is not well reproduced by both simulation packages.  The measured rate 
is 67\,$\pm$\,1 events per day, while \Fluka returns 41\,$\pm$\,3 per
day and \Geant4 returns 42.5\,$\pm$\,0.2 (Model III) 
and 44.6\,$\pm$\,0.2 (Model IV).
This discrepancy is also apparent from the multiplicity plot in
figure~\ref{fig:multiplicity} and seems to be associated with low
multiplicity events.  No explanation has been found for this
discrepancy.

\section{Conclusions}
\label{sec::conclusions}

The Borexino detector offers a unique opportunity to study cosmic
backgrounds at a depth of 3800\,m w.e. at the Gran Sasso underground
laboratories. The results are not only essential to low-energy
neutrino analyses, but are also of  
substantial interest for direct dark matter and $0\nu\beta\beta$
searches at underground facilities. Based on thermal neutron captures
in the scintillator target of Borexino, a spallation neutron yield of
$Y_{n} = (3.10 
\pm 0.11) \cdot 10^{-4}\,n/(\mu \cdot (\rm{g/cm}^{2}))$ was
determined. The lateral distance profile was measured based on
the reconstructed parent muon tracks and neutron capture vertices. An
average lateral distance of 
$\lambda=(81.5\pm2.7)$\,cm was found. The data results on neutron
yield, multiplicity and lateral distributions were compared to
Monte Carlo simulation predictions by the \Fluka and \Geant4 framework
and are largely compatible. The simulated neutron
yield of \Fluka shows a deficit of $\sim$20\,\%, 
while the result of the \Geant4 simulation is in  good agreement with
the measured value. However, both simulations should be increased as a result 
on an underprediction of $^{11}$C production.

The production rates of several cosmogenic radioisotopes in the
scintillator were determined based on a simultaneous fit to
energy and decay time distributions. Results of a corresponding
analysis performed by the KamLAND collaboration 
for the Kamioka underground laboratory \cite{Abe:2009aa} are similar to our findings.
 Moreover, Borexino rates were
compared to predictions by \Fluka and \Geant4: While there is good
agreement within their uncertainties for most 
isotopes, some cases ({$^{12}$B}, {$^{11}$C}, {$^{8}$Li} for both
codes and {$^{8}$B}, {$^{9}$Li} for \Geant4 only) show a significant
deviation between data and Monte Carlo simulation predictions.  
 
 \section*{Acknowledgements}
 
The Borexino program is made possible by funding from INFN
(Italy), NSF (USA), BMBF, DFG (OB 168/1-1), MPG, and the Garching accelerator laboratory MLL (Germany), Russian Foundation for
Basic Research (Grant 12-02-12116) (Russia), MNiSW (Poland), and
the UnivEarthS LabEx programme (ANR-11-IDEX-0005-02) (France).
We acknowledge the generous support of the Gran Sasso National
Laboratory (LNGS).

\bibliographystyle{h-physrev}
\bibliography{bxcosmogenics}

\end{document}